\documentclass[manuscript]{aastex63}
\usepackage{natbib}

\newcommand{\bl}{\textcolor[rgb]{0,0,0}}

\newcommand{\gr}{\textcolor[rgb]{0,0,0}}

\def\gtsima{$\;\buildrel > \over \sim \;$}
\def\simgt{\lower.5ex \hbox{\gtsima}}
\def\ltsima{$\;\buildrel < \over \sim \;$}
\def\simlt{\lower.5ex \hbox{\ltsima}}
\def \CHp{\ifmmode{\rm CH^+}\else{$\rm CH^+$}\fi}
\def \HH{\ifmmode{\rm H_2}\else{$\rm H_2$}\fi}
\def \Cp{\ifmmode{\rm C^+}\else{$\rm C^+$}\fi} 
\def \cc    {\ifmmode{\,{\rm cm}^{-3}}\else{$\,{\rm cm}^{-3}$}\fi}
\def \dens{\ifmmode{n_{\rm H}}\else{$n_{\rm H}$}\fi}
\def \kms   {\ifmmode{\,{\rm km}\,{\rm s}^{-1}}\else{km s$^{-1}$}\fi} 

\begin{document}

\title{Observations and analysis 
of CH$^+$ vibrational emissions from the young, carbon-rich \\
planetary nebula NGC 7027: a textbook example
of chemical pumping}

\author{David A. Neufeld}
\affiliation{Department of Physics \& Astronomy, Johns Hopkins Univ., Baltimore, MD 21218, USA}

\author{Benjamin Godard}
\affiliation{Observatoire de Paris, PSL Universit\'e, Sorbonne Universit\'e, LERMA, 75014 Paris, France}
\affiliation{Laboratoire de Physique de l’\'Ecole normale sup\'erieure, ENS, Universit\'e PSL, CNRS, Sorbonne Universit\'e, Universit\'e de Paris, 75005 Paris, France}

%\author{Miwa Goto} 
%\affiliation{Universit\"ats-Sternwarte M\"unchen, Ludwig-Maximilians-Universit\"at, 
%D-81679 M\"unchen, Germany}

\author{P.\ Bryan Changala} 
\affiliation{Center for Astrophysics $\vert$ Harvard \& Smithsonian, Cambridge, MA 02138, USA}

\author{Alexandre Faure}
\affiliation{Univ.\ Grenoble Alpes, CNRS, IPAG, F-38000 Grenoble, France}

\author{T.\ R.\ Geballe}
\affiliation{Gemini Observatory and \gr{NSF/NOIRLab}, 670 N.\ A'ohoku Place, Hilo, HI 96720, USA}

\author{Rolf G\"usten}
\affiliation{Max-Planck-Institut f\"ur Radioastronomie, Auf dem H\"ugel 69, 53121 Bonn, Germany}

\author{Karl M. Menten}
\affiliation{Max-Planck-Institut f\"ur Radioastronomie, Auf dem H\"ugel 69, 53121 Bonn, Germany}

\author{Helmut Wiesemeyer}
\affiliation{Max-Planck-Institut f\"ur Radioastronomie, Auf dem H\"ugel 69, 53121 Bonn, Germany}

\vfill\eject

\begin{abstract}

We discuss the detection of 14 rovibrational lines of CH$^+$,
obtained with the iSHELL spectrograph on NASA's Infrared Telescope Facility (IRTF) on Maunakea.
Our observations in the 3.49 -- 4.13 $\mu$m spectral region, %, conducted in 2019 July and 2020 July, 
{ obtained with a $0\farcs375$ slit width that provided a spectral resolving power
$\lambda/\Delta \lambda \sim 80,000$},
have resulted in the unequivocal detection of the $R(0) - R(3)$ and $P(1)-P(10)$ transitions
within the $v=1-0$ band of CH$^+$.  The $R$-branch transitions are anomalously weak relative
to the $P$-branch transitions, a behavior that is explained accurately 
by rovibronic calculations of the transition dipole moment 
reported in a companion paper (Changala et al.\ 2021).
Nine infrared transitions of H$_2$ were also detected in these observations, comprising the
$S(8)$, $S(9)$, $S(13)$ and $S(15)$ pure rotational \gr{lines}; the $v=1-0$ $O(4) - O(7)$
lines, and the $v=2-1$ $O(5)$ line.  We present a photodissociation region model, constrained
by the CH$^+$ and H$_2$ line fluxes that we measured, that includes a detailed
treatment of the excitation of CH$^+$ by inelastic collisions, optical pumping, and 
chemical (``formation") pumping.  \gr{The latter process is found
to dominate the excitation of the observed rovibrational lines of CH$^+$, and the
model is remarkably successful in explaining both the absolute and relative
strengths of the CH$^+$ and H$_2$ lines.}

\end{abstract}


\section{Introduction}

Planetary nebulae present an unusual astrophysical environment in which a stratified 
shell of gas is irradiated by a central star with an effective temperature that may exceed 
200,000~K.  While the inner edge of such nebulae may be highly ionized and exhibit line emission
from ions with appearance potentials in excess of 100 eV (e.g. Ne$^{5+}$), the gas 
temperature and degree of ionization drop with increasing distance from the star and even 
molecular gas can exist within the outer parts of the shell.
Molecules have been studied extensively in the young, carbon-rich planetary nebula NGC 7027, 
where roughly a dozen molecular species have been detected (e.g.\ Hasegawa \& Kwok 2001; Zhang et al.\ 2008).
These molecules have been observed
primarily through their pure rotational emissions, although infrared rovibrational emissions 
from H$_2$ have been studied over several decades (e.g. Beckwith et al. 1980; Smith et al. 1981; Cox et al. 1997, 2002).  

In the past year, rovibrational emissions
from two additional molecules have been detected towards NGC 7027 
(Neufeld et al.\ 2020; hereafter Paper I): HeH$^+$ and CH$^+$.  
Following the first astrophysical detection of HeH$^+$ by means of SOFIA/GREAT 
observations of its far-infrared $J=1-0$ pure rotational transition (G\"usten et al. 2019), 
the $v=1-0$ $P(1)$ and $P(2)$ transitions of HeH$^+$ were detected in targeted observations 
using the iSHELL spectrometer on NASA's Infrared Telescope Facility (IRTF).  These 
observations, which covered a substantial portion of the infrared $L$-band, led to the 
serendipitous discovery of a series of emission lines, spaced nearly equally
in frequency, that were identified as the $v=1-0$ $R(0) - R(3)$ and $P(1) - P(5)$ transitions 
of CH$^+$.  CH$^+$ is also a molecule that had previously been { studied} in NGC 7027 through
far-infrared observations of its pure rotational lines (Cernicharo et al. 1997; { Herpin et al.\ 2002}; Wesson et al.\ 2010).
Thus far, NGC 7027 is the only astrophysical source from which the { infrared} CH$^+$ \gr{rovibrational} lines
have been detected.

In this paper, we report the detection of four additional rovibrational emission lines 
of CH$^+$, obtained again by means of IRTF/iSHELL observations of NGC 7027; they are
the $v=1-0$ $P(7) - P(10)$ transitions.  The observations and data reduction are described
in Section 2, and the results presented in Section 3.  In Section 4, we describe a 
comprehensive model for the formation, destruction and excitation of CH$^+$, and compare
its predictions with the observed CH$^+$ \gr{$v=1-0$} line fluxes.  This model rests upon new calculations, presented in a companion paper (Changala et al. 2021; hereafter C21), 
of the spontaneous radiative rates for CH$^+$ rovibrational transitions within the ground 
electronic state and for the $A^1\Pi -X^1\Sigma^+$ band.  These calculations explain why  
the $R$-branch lines detected in the \gr{$v=1-0$} band of  CH$^+$ are anomalously 
weak compared to the $P$-branch.
In Section 5, we compare the model predictions with the observations.
A brief summary follows in Section 6.

\section{Observations and data reduction}

\begin{figure}
\includegraphics[scale=0.8,angle=-0]{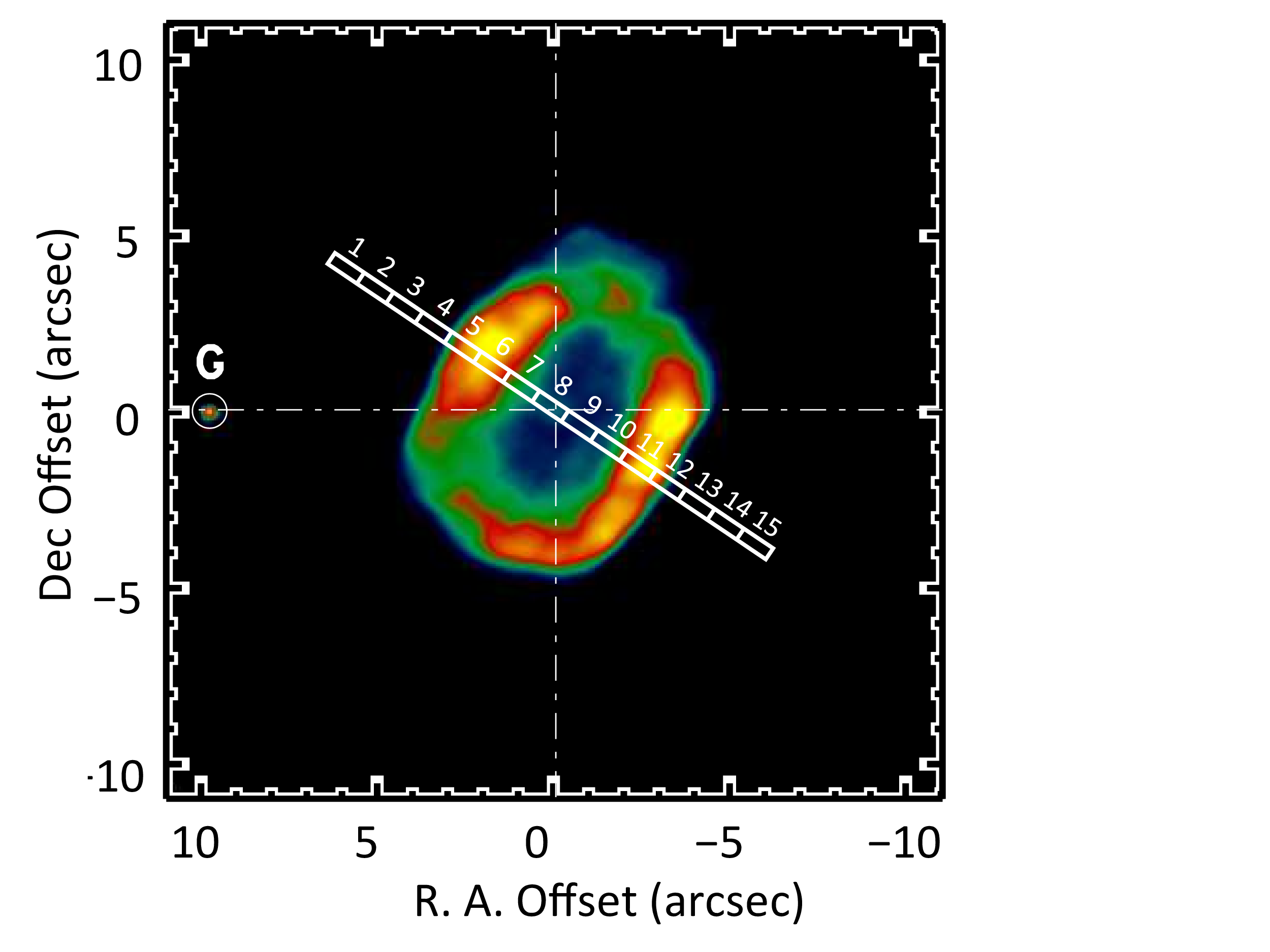}
\caption{$K$-band image of NGC 7027, obtained with the slit viewing camera.  White boxes indicate the aperture extraction regions adopted, with the numbering system used in the text.  The R.A.\ and Dec.\ offsets are given relative to the central star at $\alpha = \rm 21^h\,07^m\,1\fs 793, \delta=42\degr\,14^{\prime\prime}\,9\farcs 79$ (J2000).  ``G" indicates the star used for guiding.}
\end{figure} 

The new observations of CH$^+$
reported here were performed at the IRTF on 2020 July 08 UT,
using the iSHELL spectrograph (Rayner et al. 2016) in its {\tt Lp3} grating setting 
to cover the 3.817 to 4.155 \,$\mu$m spectral region.\footnote{These observations were 
followed by observations on 2020 July 11, 13, 14, and 15 with 
the {\tt L2}, {\tt L1}, {\tt M1}, and {\tt M2} settings, 
which -- together with the {\tt Lp1} and {\tt Lp2} observations reported in Paper I -- will provide
a complete $L$- and $M$-band line survey of NGC 7027.  Here, we focus on the 
{\tt Lp1}, {\tt Lp2} and {\tt Lp3} observations of CH$^+$ rovibrational emissions, while the
full survey will be presented in  a future paper.}
The observing  procedure was identical to that adopted for the previous observations 
and described in Paper I.
Figure 1, reproduced from Paper I, shows the slit position on a $K$-band image 
of NGC 7027 that was obtained with the slit viewing camera { and is dominated by emission 
from warm dust}.  The offsets shown here are 
relative to the position of the central star determined on the slit viewing camera: 
$\alpha = \rm 21^h\,07^m\,1\fs 793, \delta=42\degr\,14^{\prime\prime}\,9\farcs 79$ (J2000).

The slit, of length $15^{\prime\prime}$, was 
oriented at position angle 59\degr$\,$ East of North.   
{ We selected a slit width of 0\farcs375, which
provides a spectral resolving power
$\lambda/\Delta \lambda \sim 80,000$.}
The $K=11.3$~mag star 
2MASSJ~21070267+4214099, marked with a ``G'' in Figure~1, was used as 
the guide star and kept on the same pixel of the detector throughout the integration.
A reference position, located 16\arcsec~east of the {center of the} nebula, was
used to record the blank sky emission every 5 minutes while keeping the guide star 
inside the camera field-of-view.  

The calibration procedures involved the acquisition of flat fields prior to 
every change in telescope pointing, the use of sky lines for wavelength calibration,
and observations of the early-type (A0V) standard star HR\,7001 (Vega) for flux calibration.
The latter were performed at a similar airmass to that of NGC 7027, and were carried out 
with both the 0\farcs375 and 4\farcs0 wide slits in order that the slit-loss correction
could be correctly determined for the unresolved standard star.

The data reduction methods were identical to those used for the 2019 data; full details
are described in Paper I and will not be repeated here.  The data reduction pipeline made
use of the program suite {\it Spextool} ver. 5.0.2 (Cushing et al. 2004) adapted for the
iSHELL data.  This includes the {\it xtellcor} code (Vacca et al. 2003) for 
dividing the science spectra by the spectra of the standard star.  The spectra were 
obtained for 15 extraction regions defined along the slit length
(Figure~1), the size and the separation of which were each 1\arcsec. 
Using three stars in the field with Gaia astrometry, 
we determined (Paper I) that the extraction region number X (see Figure 1) is 
centered at \gr{an} offset $\Delta \theta = {\rm X} - 8.1875$ arcsec from the central 
star.  Here, the extraction regions are numbered 1 through 15 from NE to SW, 
with positive values of $\Delta \theta$ referring to offsets in the SW direction.

\section{Results}

\subsection{Spectra, position-velocity diagrams and line fluxes}

In this paper, we discuss the CH$^+$ emissions detected from NGC 7027
in the 3.265 to 4.155 \,$\mu$m spectral region 
covered by the {\tt Lp1}, {\tt Lp2} and {\tt Lp3} modes of iSHELL.  In this bandpass, 
14 rovibrational lines of CH$^+$ were detected unequivocally and are listed in Table 1
along with their wavelengths.  This table includes two additional CH$^+$ transitions, the 
$v=1-0$ $R(4)$ line, which is tentatively detected; and the $v=1-0$ $R(8)$ line, for which we
obtain an upper limit on the line flux.  The intermediate $R(5) - R(7)$ lines fall at
wavelengths that are inaccessible due to atmospheric absorption.
\gr{In addition}, three rovibrational lines and two 
pure rotational lines of H$_2$ were detected in wavelength regions
covered by the {\tt Lp1}, {\tt Lp2} and {\tt Lp3} modes of iSHELL, and two additional 
rovibrational lines and two pure rotational lines were detected using 
the {\tt L2} and {\tt M1} modes.  The H$_2$ lines that we detected are listed in Table 2.

\begin{deluxetable}{lccccc}
\tabletypesize{\footnotesize}
\tablecaption{Spectral lines of CH$^+$ observed toward NGC 7027}

\tablehead{
Line & Rest  & Upper state & Peak intensity in  & R.m.s. & Observed  \\
     & wavelength & energy / $k_B$  & {\it P-V} diagram$^a$ & noise & line flux$^b$ \\
& ($\mu$m) 	& (K)   & (MJy sr$^{-1}$) & (MJy sr$^{-1}$) & ($\rm 10^{-18}\,W\,m^{-2}$ )}
\startdata
CH$^+$ 1-0 $R(0)$ & 3.61463 & 3980 & 1125 & 102 & 2.35 $\pm$ 0.09 \\
CH$^+$ 1-0 $R(1)$ & 3.58115 & 4058 & 1248 & 84 & 2.18 $\pm$ 0.06 \\
CH$^+$ 1-0 $R(2)$ & 3.54960 & 4174 & 1107 & 69 & 1.47 $\pm$ 0.06 \\
CH$^+$ 1-0 $R(3)$ & 3.51993 & 4328 & 770 & 71 & 1.09 $\pm$ 0.06 \\
CH$^+$ 1-0 $R(4)$ & 3.49209 & 4520 & 267 & 86 & 0.21 $\pm$ 0.07 \\
CH$^+$ 1-0 $R(8)$ & 3.39839 & 5667 & 225 & 102 & --0.21 $\pm$ 0.09 \\

CH$^+$ 1-0 $P(1)$ & 3.68758 & 3942 & 1885 & 211 & 3.97 $\pm$ 0.17 \\
CH$^+$ 1-0 $P(2)$ & 3.72717 & 3980 & 3503 & 132 & 7.38 $\pm$ 0.11 \\
CH$^+$ 1-0 $P(3)$ & 3.76891 & 4058 & 4615 & 134 & 8.33 $\pm$ 0.11 \\
CH$^+$ 1-0 $P(4)$ & 3.81288 & 4174 & 4637 & 219 & 7.67 $\pm$ 0.15 \\
CH$^+$ 1-0 $P(5)$ & 3.85914 & 4328 & 3709 & 85 & 7.21 $\pm$ 0.07 \\
CH$^+$ 1-0 $P(6)$ & 3.90777 & 4520 & N/A & N/A & 6.45 $\pm$ 0.21 \\
CH$^+$ 1-0 $P(7)$ & 3.95885 & 4751 & 3554 & 108 & 6.52 $\pm$ 0.08 \\
CH$^+$ 1-0 $P(8)$ & 4.01249 & 5019 & 3633 & 122 & 5.04 $\pm$ 0.09 \\
CH$^+$ 1-0 $P(9)$ & 4.06876 & 5324 & 3261 & 169 & 4.20 $\pm$ 0.13 \\
CH$^+$ 1-0 $P(10)$ & 4.12777 & 5667 & 2955 & 253 & 3.46 $\pm$ 0.19 \\

\enddata
\tablenotetext{a}{Maximum intensity when binned to 1\arcsec\ with a $\rm 5\, km\,s^{-1}$ channel width}
\tablenotetext{b}{in a $0\farcs375 \times 15^{\prime\prime}$ aperture}
\end{deluxetable}

\begin{deluxetable}{lccccc}
\tabletypesize{\footnotesize}
\tablecaption{Spectral lines of H$_2$ observed toward NGC 7027}

\tablehead{
Line & Rest  & Upper state & Peak intensity in  & R.m.s. & Observed  \\
     & wavelength & energy / $k_B$  & {\it P-V} diagram$^a$ &  noise & line flux$^b$ \\
& ($\mu$m) 	& (K)   & (MJy sr$^{-1}$) & (MJy sr$^{-1}$) & ($\rm 10^{-18}\,W\,m^{-2}$ )}
\startdata
H$_2$ 0-0 $S(8)$ & 5.05312 & 8677 & 4394 & 366 & 11.60 $\pm$ 0.23 \\
H$_2$ 0-0 $S(9)$ & 4.69461 & 10261 & 6739 & 228 & 19.04 $\pm$ 0.16 \\
H$_2$ 0-0 $S(13)$ & 3.84611 & 17444 & 1079 & 114 & \phantom{1}3.27 $\pm$ 0.10 \\
H$_2$ 0-0 $S(15)$ & 3.62617 & 21411 & 537 & 92 & \phantom{1}1.32 $\pm$ 0.08 \\
H$_2$ 1-0 $O(4)$ & 3.00387 & 6471 & 5980 & 151 & 23.91 $\pm$ 0.16 \\
H$_2$ 1-0 $O(5)$ & 3.23499 & 6951 & 10183 & 53 & 41.78 $\pm$ 0.05 \\
H$_2$ 1-0 $O(6)$ & 3.50081 & 7584 & N/A & N/A & \phantom{1}9.79 $\pm$ 0.17 \\
H$_2$ 1-0 $O(7)$ & 3.80742 & 8365 & 3962 & 129 & 11.70 $\pm$ 0.11 \\
H$_2$ 2-1 $O(5)$ & 3.43787 & 12550 & 765 & 114 & \phantom{1}2.09 $\pm$ 0.11 \\

\enddata
\tablenotetext{a}{Maximum intensity when binned to 1\arcsec\ with a $\rm 5\, km\,s^{-1}$ channel width}
\tablenotetext{b}{in a $0\farcs375 \times 15^{\prime\prime}$ aperture}
\end{deluxetable}

For CH$^+$, the line wavelengths are based on
the spectroscopic parameters for $v=0$ and $v=1$ obtained by Hakalla et al.\ (2006) 
from measurements of the $A^1\Pi -X^1\Sigma^+$ band.  Domen\'ech et al. (2018) have 
reported direct (and more precise) wavelength measurements of 
four $v=1-0$ transitions obtained by means of action spectroscopy, but such measurements have 
so far been limited to transitions involving upper states with rotational quantum numbers
less than 3.  We find that the Hakalla et al. (2006)
spectroscopic parameters\footnote{The sextic distortion coefficients, $H$,  obtained in the two studies show a discrepancy that is considerably larger than would
be consistent with their stated uncertainties.  The values and uncertainties given for $H$ by Domen\'ech et al. (2018) were based upon a fit to the new wavelength determinations presented there in combination with previous measurements of the frequencies of pure rotational transitions (Yu et al.  2016); we find that the uncertainties given by Yu et al.\ (2016) for the latter measurements were likely underestimated, in agreement with conclusion reached by Cho \& Le Roy~(2016).  We note also that the sextic distortion coefficients presented Hakalla et al.\ (2006) are in much better agreement with theoretical estimates obtained by Sauer \&  Spirko (2013).} yield a better fit 
to the set of astronomical data, which include lines with upper 
states of rotational quantum number as high as 9.  For H$_2$, we adopt the line wavelengths
presented recently by Roueff et al.\ (2019).

\begin{figure}
\includegraphics[scale=0.9]{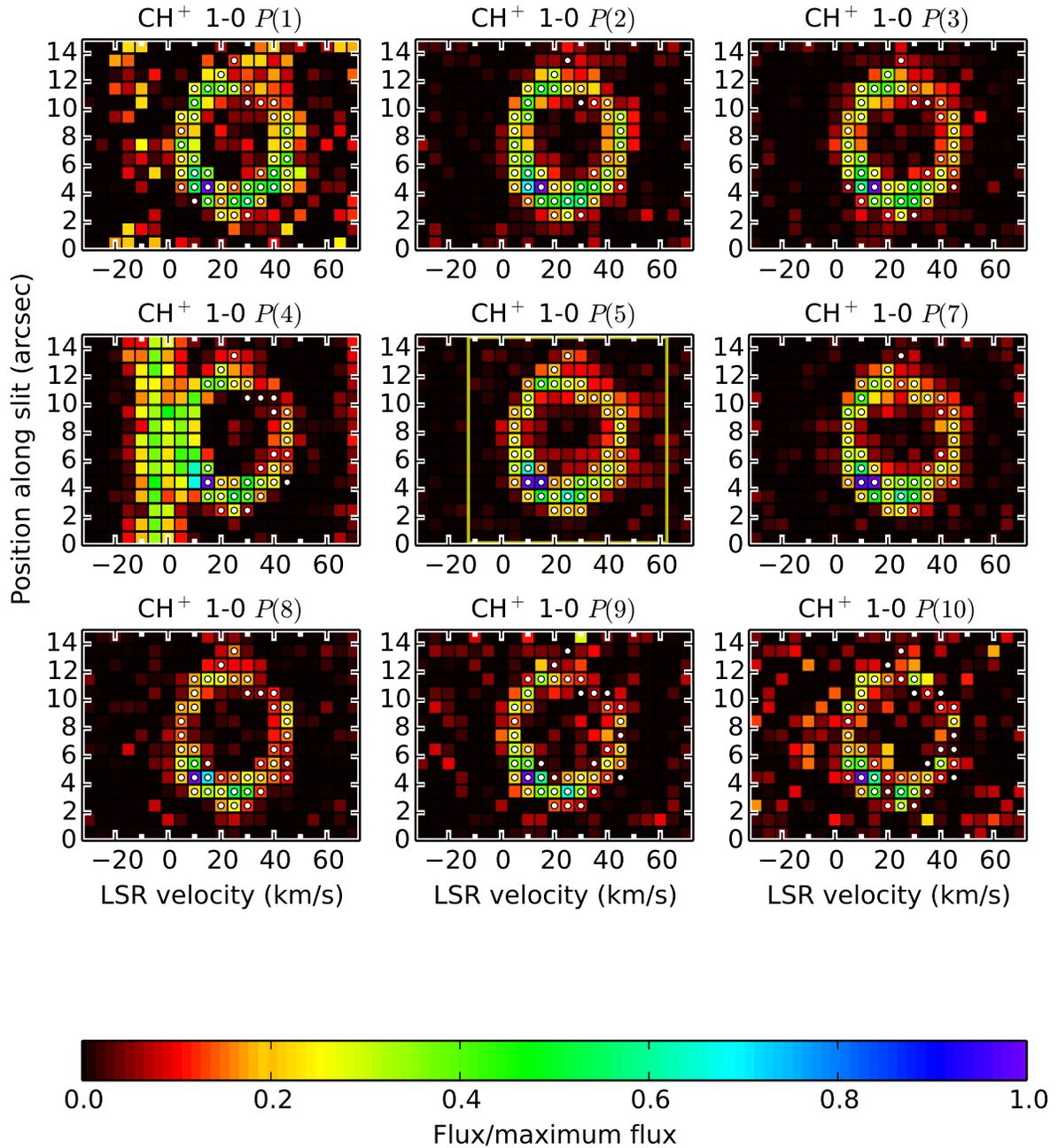}
\caption{{\it P-V} diagrams for the CH$^+$ $P$-branch lines.  The maximum intensities for each line are given in Table 1.  Yellow rectangle: region used to determine the velocity-integrated flux for the $v=1-0$ $P(5)$ line.  White dots: pixels used for flux determinations for other lines (with $v=1-0$ $P(5)$ as a template: see the text).} 
\end{figure}

\begin{figure}
\includegraphics[scale=0.9]{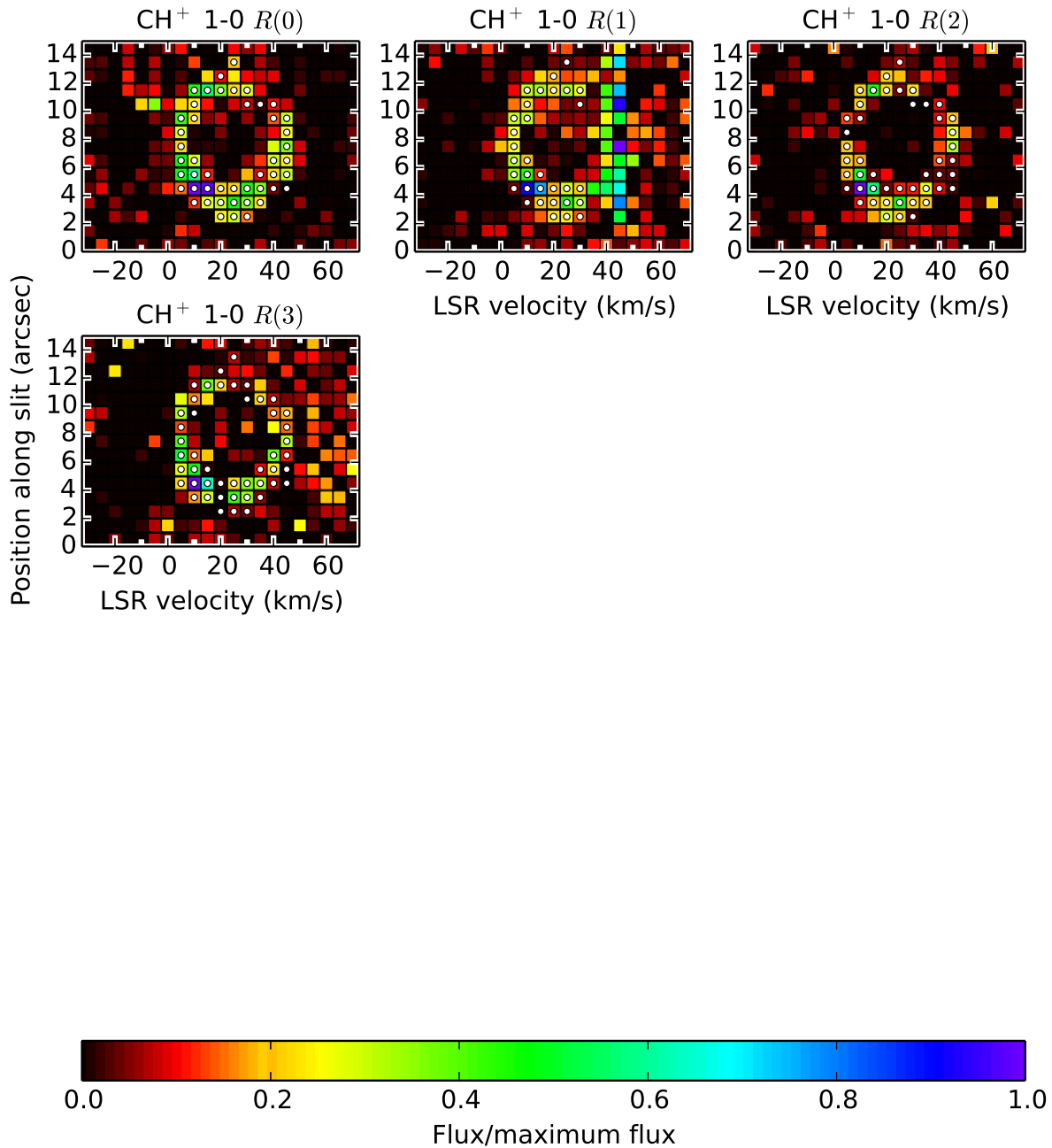}
\caption{{\it P-V} diagrams for the CH$^+$ $R$-branch lines.  The maximum intensities for each line are given in Table 1.  White dots: pixels used for flux determinations (with $v=1-0$ $P(5)$ as a template: see the text).} 
\end{figure}

\begin{figure}
\includegraphics[scale=0.9]{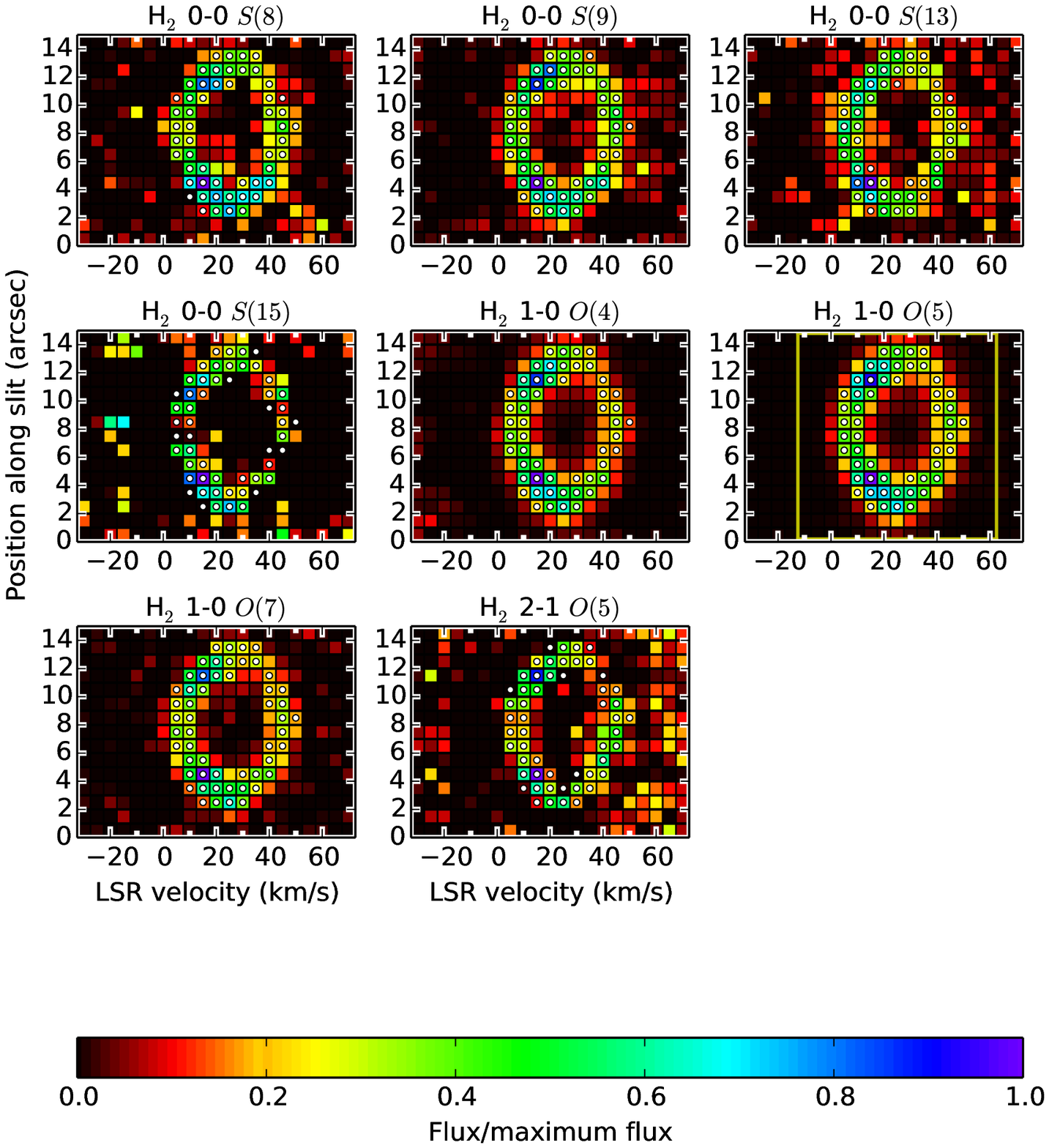}
\caption{{\it P-V} diagrams for the H$_2$ lines.  The maximum intensities for each line are given in Table 2.  Yellow rectangle: region used to determine the velocity-integrated flux for the $v=1-0$ $O(5)$ line.  White dots: pixels used for flux determinations for other lines (with $v=1-0$ $O(5)$ as a template: see the text).}
\end{figure}

In Figures 2 -- 4, we present position-velocity {\it P-V} diagrams 
for all the unequivocally detected lines in Tables 1 and 2, with the exception of the CH$^+$
$v=1-0$~$P(6)$ and H$_2$ $v=1-0$~$O(6)$ lines.  These two lines are
severely blended with strong hydrogen recombination lines and required the special 
treatment discussed in Appendix A.  In Figures 2 -- 4, spatial position is represented
along the vertical axis and Doppler shift along the horizontal.  The former is the distance
along the slit in arcsec, binned to 1 arcsec, and the latter is the velocity
relative to the Local Standard of Rest (LSR), binned to $5\,\rm km \, s^{-1}$.  Each pixel 
is colored according to the intensity (color scale at bottom), normalized (separately for
each transition) relative to the maximum flux in any pixel.  The maximum intensities
are listed in Tables 1 and 2.  The {\it P-V} diagrams presented in Figures 2 -- 4 show the intensities
measured after subtraction of a zeroth order continuum baseline at each position.
In Figures 5 and 6, we show the continuum-subtracted spectra for all the unblended
and equivocally-detected lines 
listed in Tables 1 and 2, obtained from a sum over the full slit.

\begin{figure}
\includegraphics[scale=0.8]{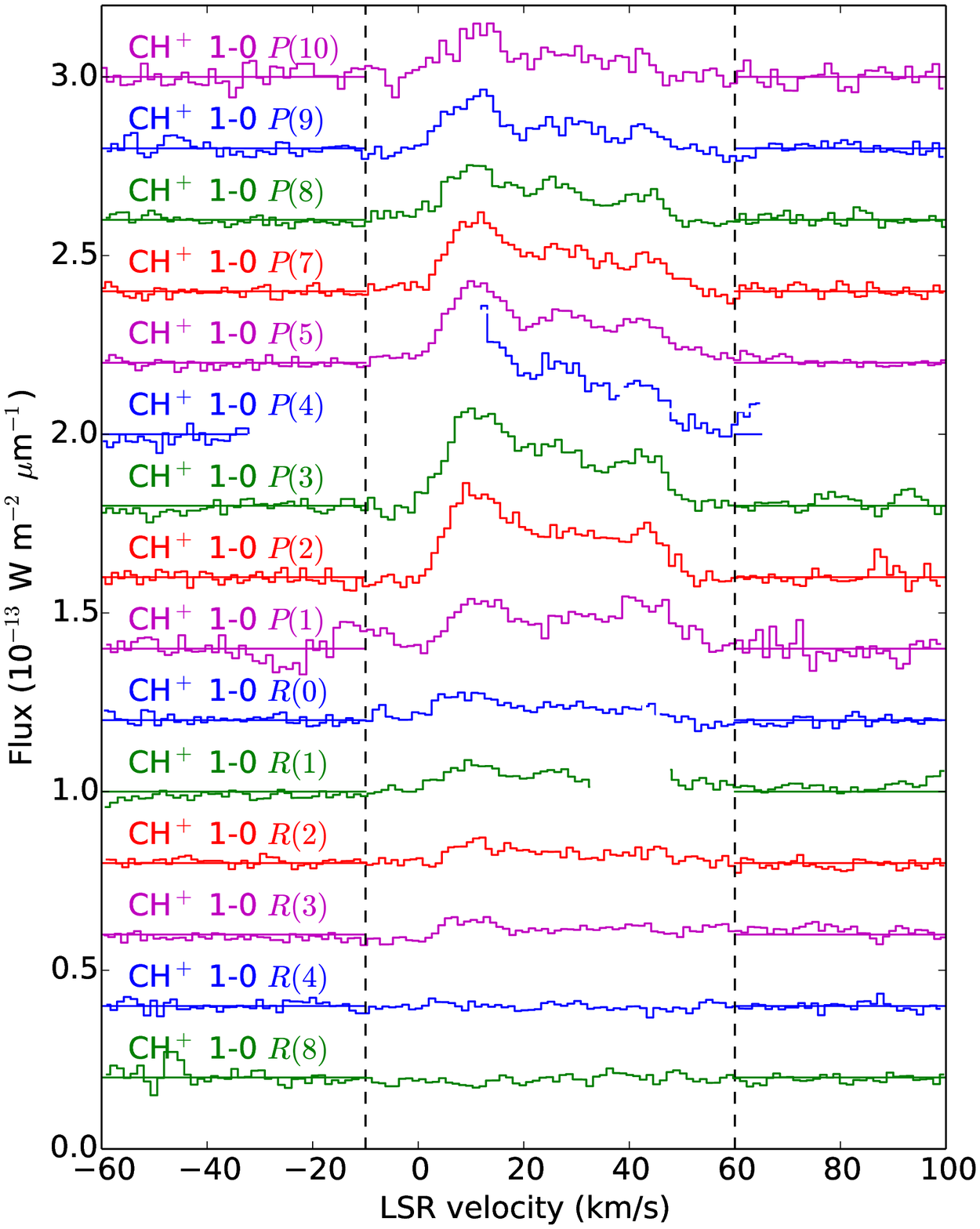}
\caption{CH$^+$ line spectra, integrated along the slit} 
\end{figure}

\begin{figure}
\includegraphics[scale=0.8]{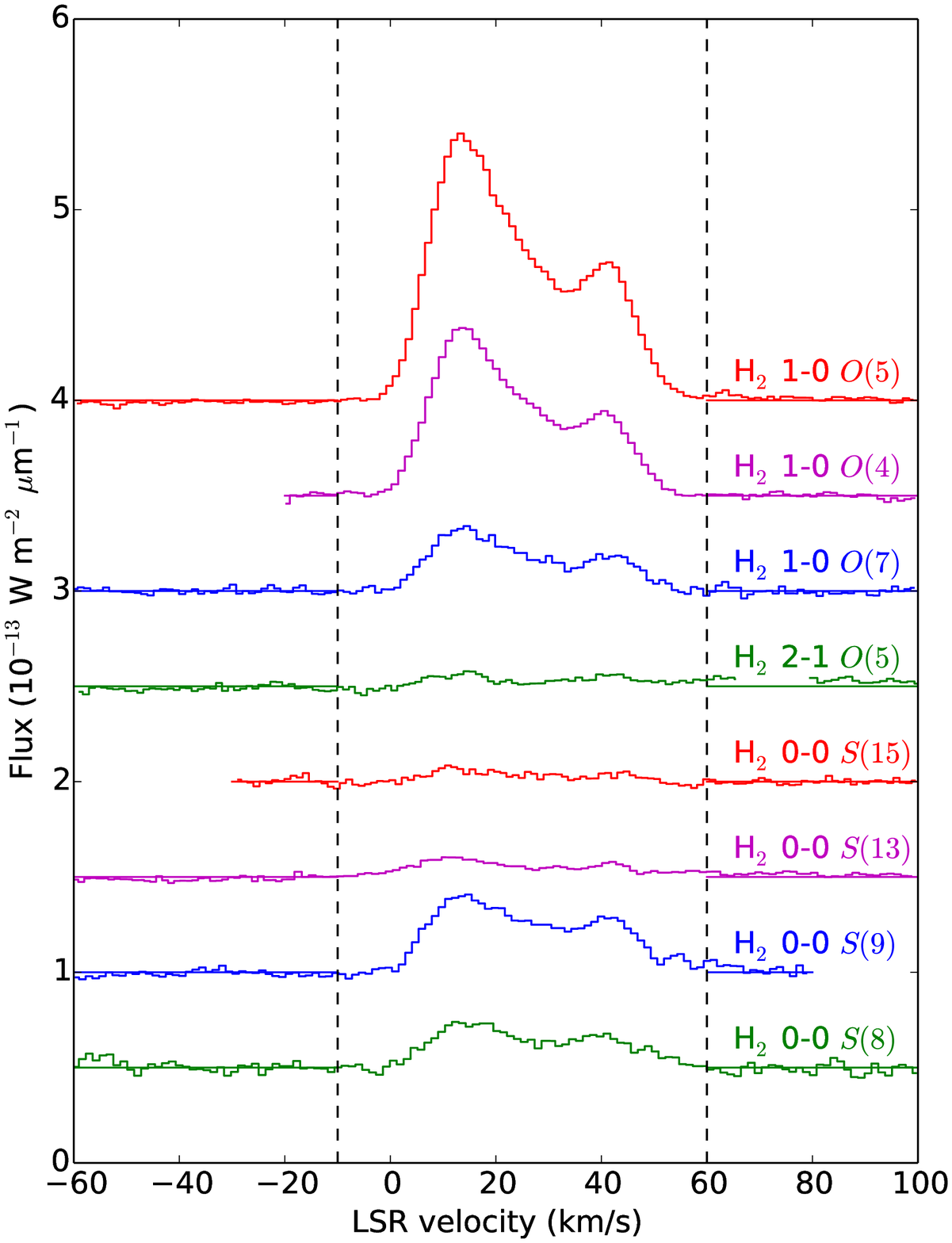}
\caption{{H$_2$ line spectra, integrated along the slit} } 
\end{figure}

For all the lines shown in Figures 2 -- 4, the {\it P-V} diagram shows a ring-like morphology, 
precisely the behavior expected when an expanding shell is observed using long-slit 
spectroscopy along the diameter.  In the case of the strong CH$^+$ $v=1-0$ $P(5)$ line, we
obtained an estimate of the total velocity-integrated line flux by summing the flux measured within the
yellow rectangle shown in the {\it P-V} diagram.  This corresponds to an integration over 
LSR velocities ranging from --12.5 to $62.5\,\rm km \, s^{-1}$, and yields the value
given in Table 1.  Here, the line is detected at a signal-to-noise ratio of
60.  

For weaker lines, however, the ring-like morphology in the {\it P-V} diagrams
suggests that a more robust line detection strategy comes \gr{from} considering the full
two-dimensional data set prior to summing the spectra at different positions.
For the CH$^+$ $v=1-0$ P(5) line, the signal-to-noise ratio can be optimized
if we include only the brightest $N$ pixels within the yellow rectangle, where
the optimal value of $N$ is found to be 47.  These pixels, marked with a white dot 
and representing 20\% of all the pixels inside the yellow rectangle, define
an annulus of strong emission for which the summed flux accounts for 80\% of the total.
For these pixels alone, the line is detected at a signal-to-noise ratio of 110.

To obtain the most robust 
determination of the fluxes for most other CH$^+$ lines in Table 1, therefore, we took a sum over the
pixels marked with a white \gr{dot}, and then divided by \gr{0.80} to account for the weak emission
lying outside the annulus of strong emission.  In the case of the CH$^+$ $v=1-0$ $P(4)$
and $R(1)$ lines, however, the {\it P-V} diagrams show clear evidence for imperfectly subtracted 
sky emission covering part of the relevant velocity range; for these lines, part of 
the 46-pixel annulus was therefore \gr{excluded} in determining their ratios to the 
CH$^+$ $v=1-0$ $P(5)$ line.   Values of the estimated line fluxes are given in Table 1, 
together with their \gr{$1\,\sigma$} statistical uncertainties.  For the subset of lines
detected in the 2019 observations, the fluxes differ somewhat from those tabulated 
in Paper I, owing to the more sophisticated analysis adopted here, and the systematic uncertainty 
estimates are smaller.  A similar procedure was adopted for the H$_2$ lines shown in Figure 4, 
using the H$_2$~$v=1-0$~$O(5)$ line as a template in place of CH$^+$~$v=1-0$~$P(5)$. 

One surprising behavior is immediately apparent from the CH$^+$ line fluxes presented in Table 1 \gr{and the spectra shown in Figure 5}:
the line fluxes are much larger for the $P$-branch lines than for the $R$-branch lines.  
Considering line pairs that originate in the same upper state, we find line flux ratios for 
$R(0)/P(2)$, $R(1)/P(3)$, $R(2)/P(4)$, $R(3)/P(5)$, and $R(4)/P(6)$ that are up
to a factor 30 smaller than those expected if the relative rovibronic matrix elements 
were simply proportional to the rotational H\"onl-London factors.  In the companion paper
C21, this anomaly is explained beautifully by
rovibronic calculations of the $J$-dependent 
transition dipole moment for the $v=1-0$ band.  
The latter is found to be unusually small and shows a significant fractional change
from one observed transition to the next; indeed, it passes through zero and switches sign 
in the vicinity of the $R(8)$ transition, and is smaller for the $R$-branch transitions than
for the $P$-branch transitions.  The results presented by C21 also
include the spontaneous radiative rates for rovibrational bands involving states up to 
$v=4$ in the ground electronic state, $X^1\Sigma^+$, and for electronic transitions in the 
$A^1\Pi -X^1\Sigma^+$ 
band.  Although vibrational bands other than $v=1-0$ have not yet been 
observed, they may play an important role in the radiative cascade that populates the
upper states of the transitions listed in Table 1, as will be discussed in Section 4.

\subsection{Rotational diagrams}

Using the spontaneous radiative decay rates presented by C21, we 
obtained the rotational diagram shown in Figure 7 (upper panel) for CH$^+$.  Here, 
log$_{10}[N(J_U)/g_U]$ is plotted as function of $E_U/k_B$, where $N(J_U)$ is the slit-averaged
column density for the $(v,J)=(1,J_U)$ upper state, $g_U = 2J_U+1$ is the degeneracy,
and $E_U$ is the energy.  The column densities $N_U(J_U)$ were computed by assuming that the
line emission is optically-thin, a good assumption because
the transition dipole moment for the $v=1-0$ band is so small.  Red points were obtained from
the observed $R$-branch line fluxes, and blue points from the $P$-branch line fluxes.
The black curve is the best fit obtained with the column densities taken as the
sum of two components with different excitation temperatures, $T_1 \sim 115$~K and $T_2 \sim 770$~K.

\begin{figure}
\includegraphics[scale=0.6]{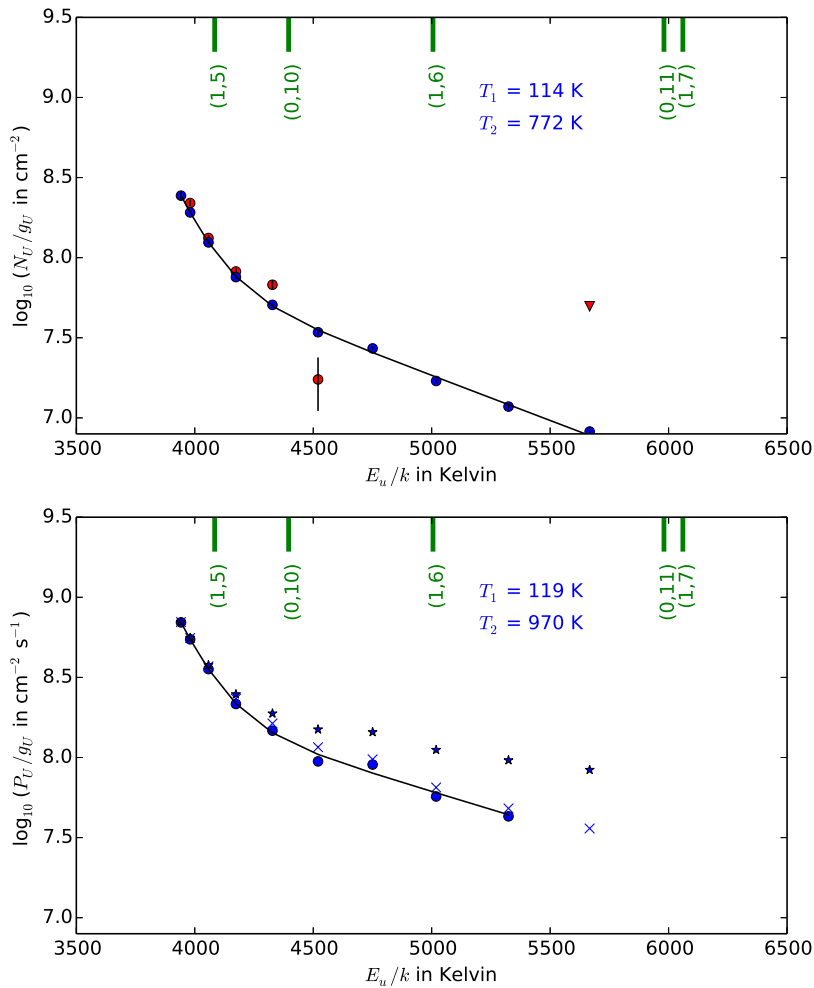}
\caption{CH$^+$ rotational diagram.  Upper panel: column densities, with a 
two-component fit (black curve).  Red points were obtained from
the observed $R$-branch line fluxes, and blue points from the $P$-branch line fluxes.
Lower panel: pumping rates under three assumptions (see the text for details).  
The filled blue circles in the lower panel represent the actual entry rates to 
$v=1$, for which we obtained a two-component fit (black curve).
\gr{The green bars at the top of each panel, which are labeled with the $(v,J)$ of 
the reactant H$_2$ molecule, indicate the energies $E_U$ above which a given reaction 
channel becomes { endoergic} (see the text for details)}.
} 
\end{figure}

In the lower panel, the column densities are multiplied by the radiative decay rates 
of each state to determine the total required rate of population, $P_U$, in equilibrium.
Results shown by the blue crosses include only radiative decay to the ground 
vibrational state: the population rates are therefore given here by $N(J_U) A_{10}(J_U)$, 
where $A_{10}(J_U)$ 
is the total spontaneous radiative rate from $(v,J)=(1,J_U) \rightarrow (0,J_U \pm 1)$ 
(i.e. including both $P$- and $R$-branch transitions).  Because the vibronic transition dipole moment 
for the $v=1-0$ band shows a significant dependence on $J_U$, $A_{10}(J_U)$ cannot be taken 
as a constant; thus, the blue crosses in the lower panel show $P_U/g_U$ declining less
rapidly with $E_U$ than does $N(J_U)/g_U$ (filled blue circles in the upper panel).  The 
black stars in the lower panel show results obtained with the inclusion of radiative
transitions {\it within the $v=1$ band}: the population rates indicated by the black stars 
are given by $N(J_U) [A_{10}(J_U)+A_{11}(J_U)]$, where $A_{11}(J_U)$ is the spontaneous
radiative rate for the $(v,J) = (1,J_U) \rightarrow (1,J_U-1)$ transition.  These
points lie above the blue crosses because they now include $v=1-1$ transitions in the radiative 
loss rate; even though the $v=1-1$ transitions are at much lower frequency than the 
$v=1-0$ transitions, the dipole matrix element is much larger and $A_{11}(J_U)$ exceeds
$A_{10}(J_U)$ for $J_U \ge 8$.  
%The pumping rates represented by the black stars include
%the effects of transitions from $v=1$ states with $J > J_U$.  

The results represented by the black stars include two contributions to the rate
at which a given state is populated: the entry rate to $v=1$, and the
radiative decay rate from $(v,J) = (1,J_U+1) \rightarrow (1,J_U)$.  To determine
the former alone, we make use of the expression 
\begin{equation}
P_U = N(J_U) [A_{10}(J_U)+A_{11}(J_U)] - N(J_U+1)A_{11}(J_U+1).
\end{equation}
This population rate is represented by the filled blue circles,
and may be considered the true entry rate to $v=1$.  
As will be discussed
in Section 4 below, the population of vibrationally-excited states of CH$^+$ 
is dominated by chemical (or ``formation") pumping, in which the reaction of C$^+$ with
H$_2$ to form CH$^+$ leaves CH$^+$ in a vibrationally-excited state.  The 
filled blue circles in the lower panel therefore represent the rate of chemical 
pumping.  As in the upper panel, we fitted the blue circles representing $P_U/g_U$
by the sum of two thermal distributions at different temperatures: the best-fit values were
$T_1 \sim 120$~K and $T_2 \sim 970$~K.

The green bars at the top of each panel, which are labeled with the $(v,J)$ of 
the reactant H$_2$ molecule, indicate the energies $E_U$ above which a given reaction 
channel becomes { endoergic}.  In other words, the reaction of H$_2(v,J)$ with 
C$^+$ to form an excited state of CH$^+$ with energy $E_U$ is { exoergic} for $E_U$ 
to the left of the green bar labeled with $(v,J)$ and { endoergic} to the right.

\begin{figure}
\includegraphics[width=16 cm, trim= 0cm 0cm 0cm 0cm, clip]{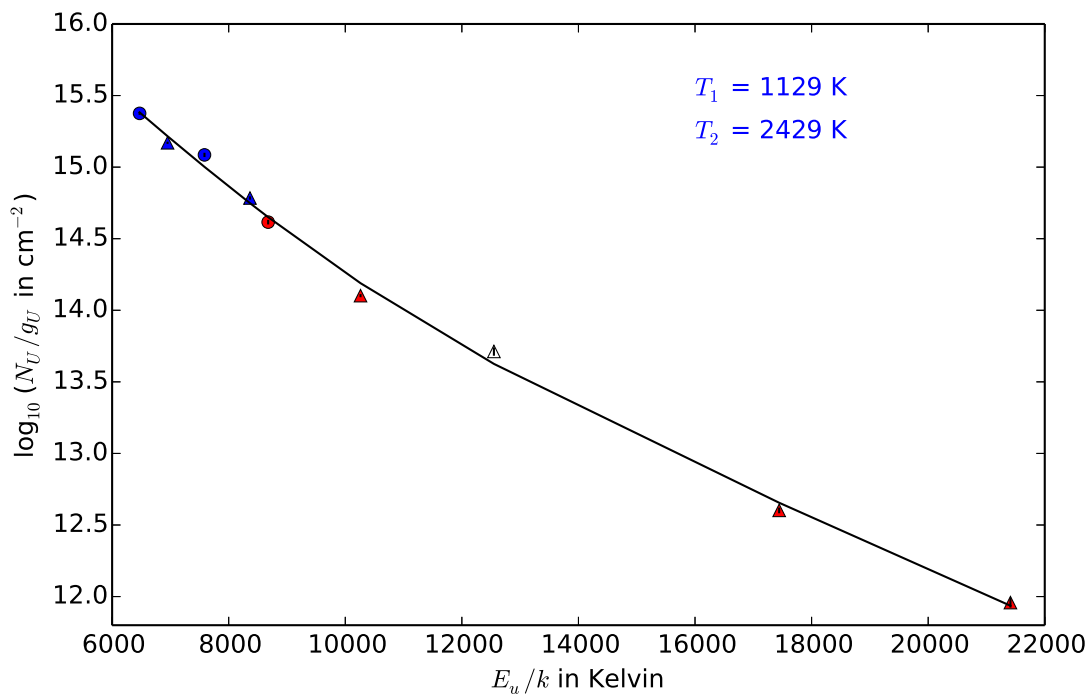}
\caption{H$_2$ rotational diagram.  Red, \gr{blue} and white symbols refer to states with $v=0$,
1 and 2 respectively; triangles represent states of ortho-H$_2$ (i.e.\ with $J_U$ odd)
and circles represent states of para-H$_2$ ($J_U$ even); 
the black line is a two-component fit. } 
\end{figure}

In Figure 8, we show the rotational diagram for the observed transitions of H$_2$.
Once again, the black curve is the best fit obtained with the column densities taken as the
sum of two components with different excitation temperatures, in this case
\gr{$T_1 \sim 1130$~K and $T_2 \sim 2430$~K.}  Red, \gr{blue} and white symbols refer to states with $v=0$,
1 and 2 respectively; triangles represent states of ortho-H$_2$ (i.e.\ with $J_U$ odd)
and circles represent states of para-H$_2$ ($J_U$ even).  Overall, a simple two-temperature
model yields a reasonably good fit to all the observed line fluxes, regardless of vibrational
state\footnote{One systematic deviation from the fit is that the column densities of 
para-H$_2$ states in  $v=1$ are slightly elevated relative to the ortho-H$_2$ states, 
implying that the ortho-to-para ratio in $v=1$ is smaller than the value of 3 
expected in local thermodynamic equilibrium.
As discussed by Sternberg \& Neufeld (1999), this behavior can be readily 
understood as being the
result of differential shielding in the optically-thick ultraviolet transitions
that pump the $v=1-0$ band of H$_2$}.

\section{\bl{Model for the excitation of \CHp}}

Our chemical and excitation model \bl{for} \CHp\ updates that described by Godard \&
Cernicharo (2013; hereafter GC13) and extends it to include predictions for the rovibrational lines. The GC13 model, which includes \CHp\ excitation by formation pumping, collisional excitation, and radiative pumping at optical and infrared wavelengths, was  successfully used to model \CHp\ pure rotational emissions detected from NGC 7027 (Cernicharo et al.\ 1997). The chemistry, excitation, and radiative transfer of \CHp\ are computed here with the latest version of the Meudon PDR code\footnote{version 1.5.4 available on the ISM platform \url{https://ism.obspm.fr}} (e.g. Le Petit et al.\ 2006) modified to treat the entire energetic structure of \CHp\ and all excitation and deexcitation processes. In this section, we describe updates to the treatment of the fundamental physical processes of relevance.

\subsection{\CHp\ energy structure and spectroscopy}

As in GC13, we include all rovibrational levels with $v \leqslant 4$ and $J \leqslant 30$ within the ground electronic state $X^1\Sigma^+$ and the first electronic state $A^1\Pi$ of \CHp. The corresponding energies are derived from the spectroscopic parameters of Hakalla et al.\ (2006) and 
%spread almost homogeneously 
\bl{range up to $E/k_B \sim 50,000$ K} (see GC13, their Figure 1). To take into account the centrifugal \bl{distortion}-induced interference effects described by C21, the spontaneous decay rates of allowed electronic and rovibrational transitions of \CHp\ are calculated using the polynomial fits to the $m$-dependent dipole moments derived by C21. This approach  \bl{differs considerably} from the prescription used in GC13, where radiative decay rates were calculated under the Born-Oppenheimer and $r$-centroid approximations using the Franck-Condon factors and the $r$-centroids of Hakalla et al.\ (2006).

With these new data \bl{in hand}, 
the infrared pumping of the rovibrational levels of the $X^1\Sigma^+$ state is computed using the escape probability formalism. In contrast, the optical pumping of the $A^1\Pi$ electronic state, followed by the \bl{subsequent} fluorescent cascade in the ground state, is treated using pumping matrix elements under the approximation that the deexcitation of any electronic level is dominated by the radiative decay to the $X^1\Sigma^+$ state.  Self-shielding processes within the electronic lines of \CHp\ at optical wavelengths are calculated using the FGK approximation (Federman et al.\ 1979).

\subsection{Collisional excitation}

Inelastic nonreactive collisional processes are included taking into account H, \HH, He, and e$^-$ as collision partners and using the rates and prescriptions described in Appendix B.  Because it is a short-lived species, \CHp\ can be highly sensitive to formation pumping mechanisms. Indeed, in hot and dense PDRs, the ion-neutral reaction 
\begin{equation} \label{Eq-react}
\Cp + \HH(v,J) \rightarrow \CHp(v',J') + {\rm H}
\end{equation}
is found to be not only the dominant pathway for the formation of \CHp\ (e.g.\ Ag{\'u}ndez et al.\ 2010) but also a major process for the excitation of its pure rotational levels (e.g. GC13, Faure et al.\ 2017). To account for this effect, we include the detailed treatment of the above state-to-state chemical reactions \bl{described in Appendix B.}
\footnote{ We have also considered the reaction $\rm C^{++} + H_2 \rightarrow CH^+ + H^+$ as a possible source of CH$^+$ formation and excitation 
within the
photoionized region that lies interior to the PDR.
Here, we adopted an upper limit on the rate coefficient presented recently by Pla{\v{s}}il et al.\ (2021)
and used the CLOUDY photoionization model (Ferland et al.\ 2017) to determine the expected $\rm C^{++}$ and $\rm H_2$ abundances.  Even if every reaction of $\rm C^{++}$ and $\rm H_2$ leads to a vibrationally-excited $\rm CH^+$ molecule, the resultant total rovibrational line flux lies at least 4 orders of magnitude below the measured value.  This possible production mechanism can be therefore be excluded robustly.}

\subsection{Geometrical model and radiative transfer}

\begin{figure*}[!ht]
\begin{center}
\includegraphics[width=15.0cm,trim = 1cm 0.0cm 1cm 0.0cm, clip,angle=0]{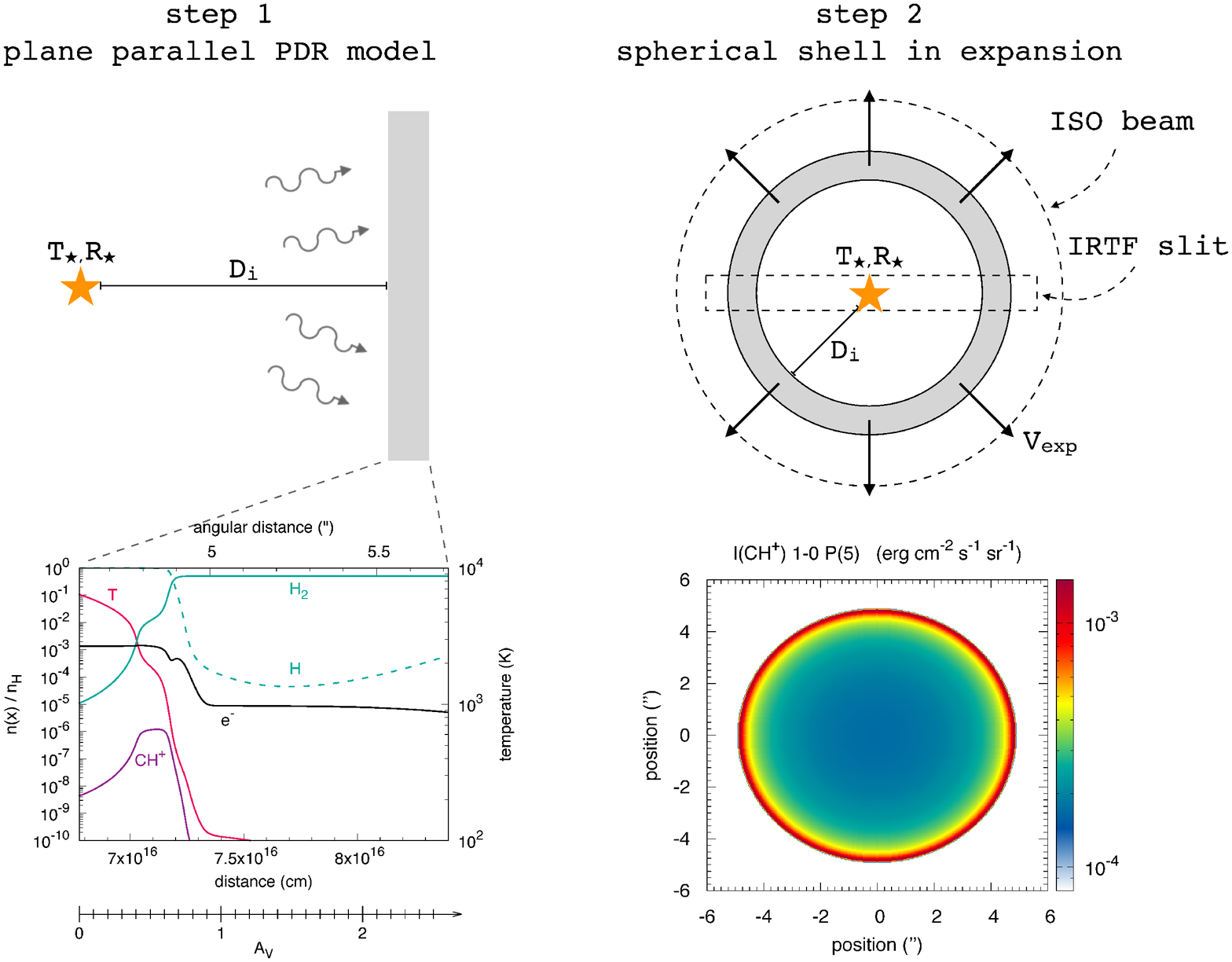}
\caption{Summary of the modeling strategy. {\it Top left :} the thermochemical and excitation profiles of a neutral slab of gas illuminated on one side by a star are calculated in a plane parallel geometry. {\it Bottom left:} kinetic temperature $T$ and abundances of H, \HH, e$^-$, and \CHp\ obtained for an isochoric PDR with a density $\dens=3\times 10^5$ \cc\ as functions of the distance to the star, starting from the ionization front.   (Results for a isochoric PDR with this density, obtained from the earlier model of GC13, are shown in their Figure 6.)
{\it Top right:} the radiative transfer within the rovibrational lines of \HH\ and \CHp\ is computed by wrapping the PDR around a sphere of radius $D_i$ expanding at a constant velocity $V_{\rm exp}$. {\it Bottom right:} example of the intensity map obtained for the $1-0$ P(5) line of \CHp.}
\label{Fig-strategy}
\end{center}
\end{figure*}

The modeling of \HH\ and \CHp\ emissions in NGC 7027 is performed \bl{in two steps as} schematized in Figure \ref{Fig-strategy}. The thermal, chemical, and excitation profiles of the neutral gas are computed with the Meudon PDR code in a plane parallel geometry, taking into account all the processes described in the previous section (bottom left panel of Figure \ref{Fig-strategy}).  As in Paper I and following Zijlstra et al.\ (2008), we
assume a distance to the source of 980~pc.    
The PDR is assumed to be illuminated from one side by a black body generated by a star with an effective temperature $T_\star=198 \, 000$ K and a radius $R_\star = 5.21 \times 10^9$ cm (Latter et al.\ 2000) located at a distance $D_i=0.022$ pc from the ionization front (see Paper I). To correctly model the self-shielding and the UV pumping of the electronic lines of \HH, the radiative transfer within the Lyman and the Werner bands of \HH\ is computed self-consistently with an exact treatment rather than with the FGK approximation. Chemical and excitation profiles are computed up to a visible extinction $A_{V\,{\rm max}}=10$.

To mimic a spherical shell in expansion at a constant radial velocity $V_{\rm exp}$, the resulting PDR is then wrapped around a sphere of radius $D_i$ (top right panel of Figure \ref{Fig-strategy}). The radiative transfer in the rovibrational lines of \HH\ and \CHp\ is solved within this geometry for 200 values of the impact parameter homogeneously spread up to the outer shell, taking into account the Doppler shift and the Doppler broadening induced by the expansion and a turbulent velocity dispersion $\sigma_{\rm turb}$. The resulting line profiles are variously integrated in order to derive intensity maps (bottom right panel of Figure \ref{Fig-strategy}), position-velocity diagrams, and line fluxes collected over the ISO beam or the IRTF slit (shown with dashed lines on the top right panel of Figure \ref{Fig-strategy}).

\subsection{Main parameters}

The PDR is modeled as an isochoric or isobaric environment neglecting the 
%necessary 
dilution of the expanding gas along the radial direction. This approximation is supported by the fact that the width of the neutral shell responsible for \HH\ and \CHp\ rovibrational emissions relative to its distance from the central star never exceeds 20\% in all our models (see, for instance, the bottom left panel of Figure \ref{Fig-strategy}). As described above, the radiation conditions are derived from the observed properties of the central star of NGC 7027. Following GC13, the carbon-rich circumstellar envelope is modeled with C, O, N, and S gas phase elemental abundances of $1.3 \times 10^{-3}$, $5.5 \times 10^{-4}$, $1.9 \times 10^{-4}$, and $7.9 \times 10^{-6}$ (Middlemass 1990). The PAHs are essentially removed by setting a PAH-to-dust mass ratio of $10^{-5}$. The density $n_{\rm H}$ (for isochoric models) and the thermal pressure $P$ (for isobaric models) are left as free parameters.

Interestingly, { despite the fact that the elemental abundances in carbon-rich evolved stars differ markedly from those in ordinary PDRs}, a first estimate of the PDR density (or pressure) can be obtained using the ISMDB Inverse Search Service\footnote{\url{https://ismdb.obspm.fr}} which performs comparisons of observational data with precomputed grids of PDR models. Applying this tool to the mean intensity of the rovibrational lines of \HH\ (Table 2) yields to a PDR with a density $n_{\rm H} \sim 10^5$ \cc\ (or a pressure $P/k_B \sim 2\times 10^8$ K \cc) illuminated by a standard interstellar radiation field scaled by a factor $G_0 \sim 10^5$, in excellent agreement with the values obtained with the model of GC13 and the UV energy density expected at a distance $D_i=0.022$ pc from the central star of NGC 7027. In the following, we therefore explore the results of our model for several values of the density and thermal pressure centered around these preliminary estimates.

\section{\bl{Comparison between the model predictions and the observations}}

\subsection{\CHp\ and \HH\ line fluxes}

\begin{figure*}[!ht]
\begin{center}
\includegraphics[width=19.0cm,trim = 1cm 2.0cm 0cm 1.0cm, clip,angle=0]{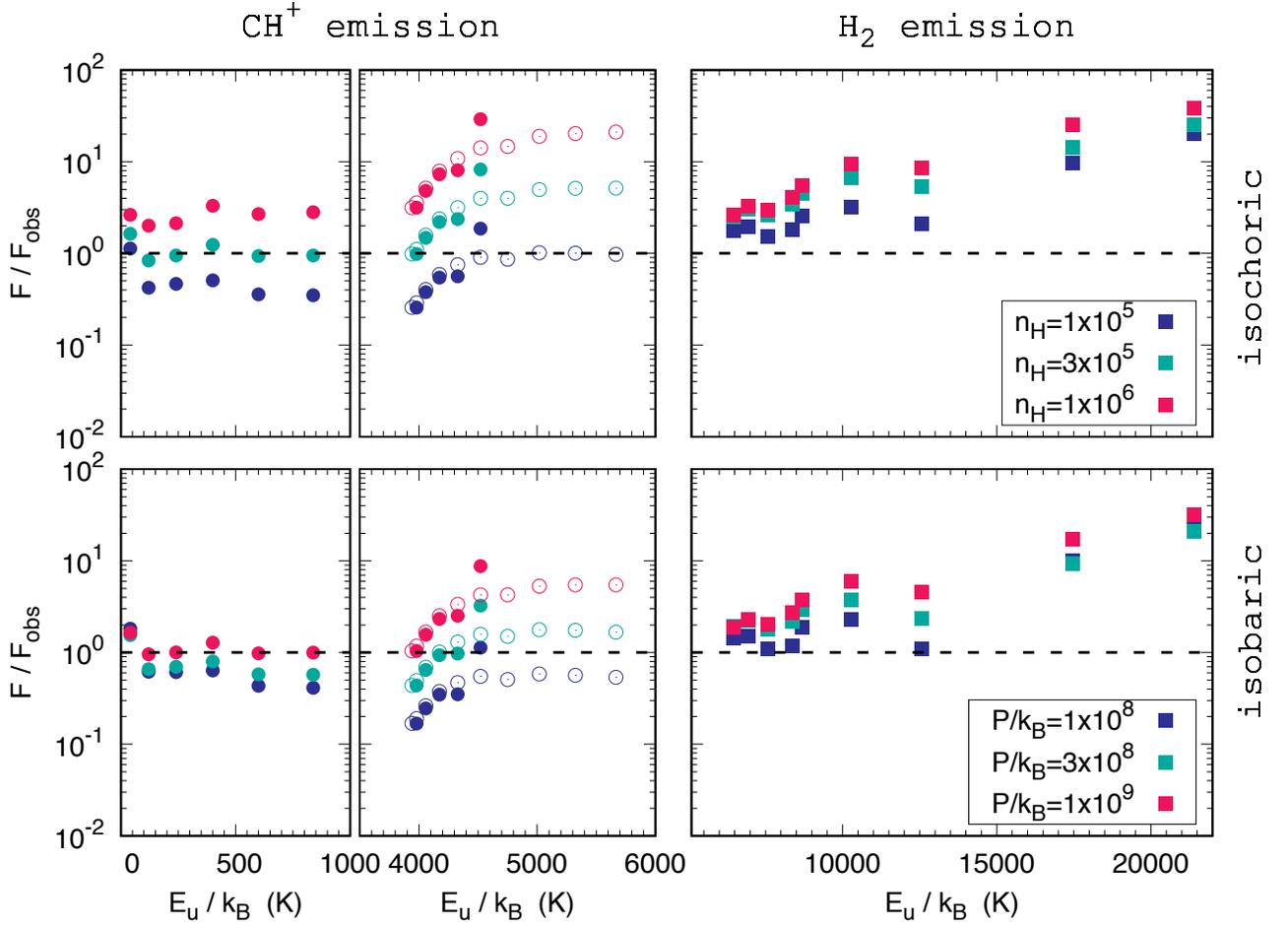}
\caption{Ratios of predicted to observed fluxes $F/F_{\rm obs}$ computed for the pure rotational lines of \CHp\ (left panel), the rovibrational lines of \CHp\ (middle panel), and the rovibrational lines of \HH\ (right panel) as functions of the energies $E_u / k_B$ of the upper levels of the transitions. The models include isochoric PDRs (top panels) with $n_{\rm H} = 10^5$, $3\times 10^5$, and $10^6$ \cc\ and isobaric PDRs (bottom panels) with $P/k_B = 10^8$, $3\times 10^8$, and $10^9$ K \cc. R-branches and P-branches are shown with filled and empty circles, respectively. Simulated fluxes are obtained by integrating the line intensities over the instrumental beam (see text for details).}
\label{Fig-fluxes}
\end{center}
\end{figure*}

A comparison between the predicted and the observed line fluxes of \CHp\ and \HH\ is shown in Figure \ref{Fig-fluxes} which displays the results obtained for isochoric and isobaric PDRs at different densities and thermal pressures, assuming a turbulent velocity dispersion $\sigma_{\rm turb}=$ 6 \kms. To match the various observational setups, the fluxes of the pure rotational lines of \CHp\ are integrated over a circular aperture with a diameter of 37\arcsec\ ($J=1-0$, observed with
{\it Herschel}/SPIRE by Wesson et al.\ 2010) or
85\arcsec\ (other pure rotational
transitions, observed with {\it ISO} by Cernicharo et al.\ 1997), while those of the rovibrational lines of \CHp\ and \HH\ are integrated over a slit of 0.375\arcsec$\times$15\arcsec\ (see Figure \ref{Fig-strategy}). 

Over the entire grid of models, the excitation of the high energy pure rotational levels of \CHp\ ($J \simgt 4$) and the excitation of all its rovibrational levels are dominated by formation pumping. In particular, non-reactive collisional excitation of the $v=1$ levels is found to be at least one order of magnitude less efficient than formation pumping even if the scaling applied for collisional transition between vibrational levels (i.e. the $a$ coefficient in Eq. \ref{Eq-proba-dist}) is set to 1. Similarly, and in line with the results of GC13, the radiative pumping of the first electronic state of \CHp\ has a weak contribution. Because the stellar radiation field follows a Planck law at high temperature, the energy density at optical wavelengths is negligible compared to the UV energy density and at least 100 times too small to make the radiative pumping a dominant source of excitation of \CHp.

\bl{The fraction of the incident UV flux that is reprocessed into rovibrational lines of H$_2$ depends on the density of the PDR.  Because the excitation of the lower rotational levels of H$_2$ varies non-linearly with the density and temperature, the self-shielding of H$_2$ is less efficient at higher densities; hence the radiative pumping of the rovibrational lines of H$_2$ extends over larger column densities when the density is higher.}

Since both the formation of \CHp\ and its excitation result from state-to-state chemistry involving molecular hydrogen, the dependence of \CHp\ emission on the gas density is even stronger. As previously shown by GC13, we find that an isochoric PDR with a density $\dens = 3 \times 10^5$ \cc\ almost perfectly reproduces the observed fluxes of all the pure rotational lines of \CHp, with discrepancies smaller than a factor of 1.5. However, such a model overestimates the emissions of several rovibrational lines of \CHp\ by a factor of 4 to 6 and those of the rovibrational lines of \HH\ by a factor of 3 to $\sim 30$. In contrast, we find that an isobaric PDR with a thermal pressure $P/k_B = 3 \times 10^8$ K \cc\ is able to reproduce simultaneously the emissions of most of the rovibrational lines of \CHp\ and \HH\ within a factor of 2 to 3, with the exception of the $0-0$ S(15) and S(13) lines of \HH. Interestingly, the first two \bl{pure} rotational transitions of \CHp\ are optically-thick and thus their line strengths depend strongly on the turbulent velocity dispersion $\sigma_{\rm turb}$. Reproducing the observed intensities of these two lines with the optimal isochoric or isobaric models requires $6 \kms\ \leqslant \sigma_{\rm turb} \leqslant 8$ \kms\ in good agreement with the velocity dispersion derived from the wings of \CHp\ and \HH\ spectra (see Figures 5 and 6).

\subsection{Position-velocity diagrams}

\begin{figure}[!ht]
\begin{center}
\includegraphics[width=9cm,trim = 0cm 6cm 0cm 0cm, clip,angle=0]{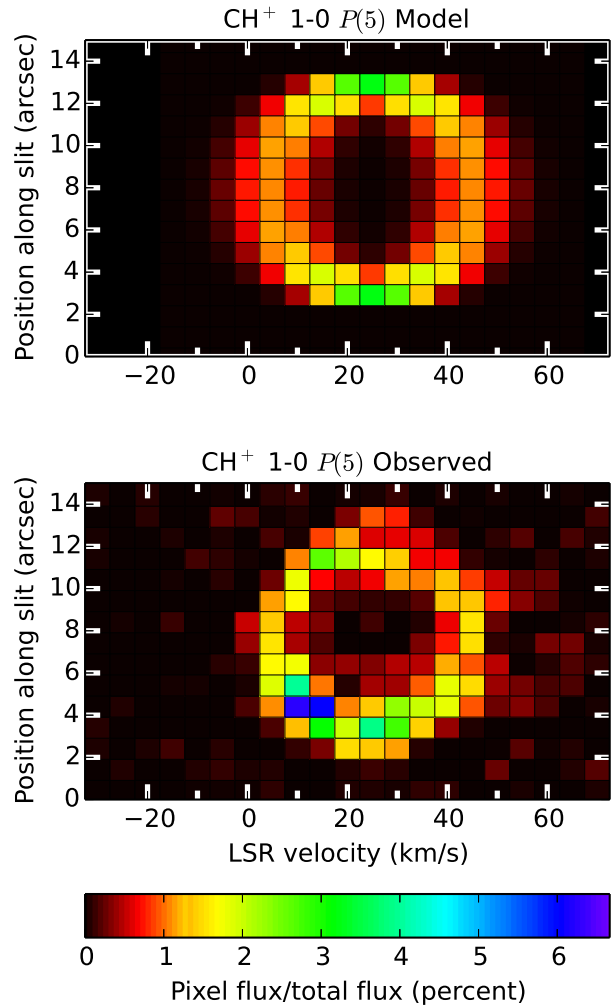}
\caption{Top panel: Specific intensity of the $1-0$ P(5) line of \CHp\ computed using an isochoric PDR model with $\dens=3\times 10^5$ \cc\ wrapped around a sphere expanding at a velocity $V_{\rm exp}=20$ \kms\ as a function of the velocity and the position along a slit identical to the observational setup. 
Intensities are averaged over a spatial bin of 1\arcsec\ and a spectral bin of 5 \kms\ as in Figures 2 – 4.
\bl{The
source was assumed to have a centroid $v_{\rm LSR}$ of 25~$\rm km\, \rm s^{-1}$ 
with its center at +8\arcsec\ . 
Bottom panel: observed PV diagram for \CHp\ $1-0$ P(5)}}
\label{Fig-pos-vel-diagram}
\end{center}
\end{figure}

The position-velocity diagram \bl{for} the $1-0$ $P(5)$ line of \CHp\ predicted along the observational slit is shown in Figure \ref{Fig-pos-vel-diagram}. The optimal isochoric (or isobaric) model explains remarkably well the shape and both the spatial and spectral extents of the ring-like profile described in Section 3, \bl{provided} that the spherical shell expands at a velocity $V_{\rm exp}=20$ \kms. Unsurprisingly, the model also predicts strong emission at the systemic velocity of the source. This is a direct consequence of the limb brightening effect and of the simplicity of the modeling. In the single plane parallel PDRs modeled here, the size $d$ of the region responsible for the emission of \CHp\ and \HH\ is small compared to the distance of the neutral cloud to the central star $D_i$. When wrapped in a spherical geometry, the limb brightening effect, i.e.\ the \bl{enhanced} emission of the border of the sphere compared to the center, scales as
\begin{equation}
\sqrt{\frac{2D_i}{d} + 1},
\end{equation}
leading to strong variations of the intensity both in space (Figure \ref{Fig-strategy}) and in velocity (Figure \ref{Fig-pos-vel-diagram}). This result strongly contrasts with the observations which indicate that the emission of \CHp\ and \HH\ is \bl{less strongly limb brightened} (see Figures 2 -- 4). Since the excitation conditions of \CHp\ and \HH\ cannot be explained by a medium at low density ($\leqslant 10^5$ \cc), which would naturally \bl{yield a thicker shell with less limb brightening},
this discrepancy implies that the geometrical model is too simplistic. 
A second, related discrepancy concerns the relative spatial extents of the different
lines that were observed.  While we detected no measurable differences between
the spatial extents of the various CH$^+$ $v=1-0$ transitions, in agreement with the
model predictions, the line emissions from H$_2$ have a slightly greater spatial extent  than those from CH$^+$ (as is evident from a comparison of Figures 2 and 4).  The model, however, predicts a $P-V$ diagram for the H$_2$ line that is indistinguishable from that observed for CH$^+$ (Figure 11, top panel).  
These discrepancies suggest that the neutral shell of NGC 7027 should \bl{ultimately} be 
modeled in a non-spherical and inhomogeneous geometry following 
the complex \bl{three-dimensional structure} of the nebula derived by Cox et al.\ (2002), for example.  \bl{Such an analysis is beyond the scope of the present work.}

\subsection{ CH$^+$ rovibrational emissions as a tracer of warm, dense UV-irradiated gas}

{ The models presented here suggest that CH$^+$ rovibrational emissions are a tracer
of warm ($T \sim 1000\, \rm K$), dense ($n_{\rm H} \sim 3 \times 10^5\,\rm cm^{-3}$)  UV-irradiated gas.  The critical pathway for the excitation of these emissions
is the reaction of C$^+$, produced by photoionization, and excited H$_2$ in states with $E/k_B \simgt 8000$~K.  In UV-irradiated gas, 
excited H$_2$ is produced both 
by radiative pumping through the Lyman and Werner bands and by collisional excitation in the warm UV-heated gas; {at densities above $\sim 10^5$ \cc (Sternberg \& Dalgarno 1989), the latter process is typically dominant in populating the lower energy ($E/k_B \le 15,000$~K) excited states that are most important for CH$^+$ formation.}
Dense PDRs (unassociated with evolved stars) and UV- and self-irradiated shock waves (e.g. Godard et al.\ 2019, Lehmann et al.\ 2020) are other environments in which 
CH$^+$ rovibrational emissions are potentially detectable; the required conditions are especially prevalent in starburst galaxies (e.g. Falgarone et al.\ 2017).}

\clearpage
\section{Summary}

1) Observations in the 3.49 -- 4.13 $\mu$m spectral region, conducted with the iSHELL spectrograph on NASA's Infrared Telescope Facility (IRTF) on Maunakea, have resulted in the unequivocal detection of the $R(0) - R(3)$ and $P(1)-P(10)$ transitions within the $v=1-0$ band of CH$^+$.  Nine infrared 
transitions of H$_2$ were also detected in these observations, comprising the
$S(8)$, $S(9)$, $S(13)$ and $S(15)$ pure rotational lines; the $v=1-0$ $O(4) - O(7)$
lines; and the $v=2-1$ $O(5)$ line.  

2) The $R$-branch transitions are anomalously weak relative
to the $P$-branch transitions, a behavior that is explained accurately 
by rovibronic calculations of the transition dipole moment 
reported in a companion paper (Changala et al.\ 2021).

3) We presented a photodissociation model that includes a detailed 
treatment of the excitation of CH$^+$ by inelastic collisions, optical pumping, and 
chemical (``formation") pumping.  

4) Chemical pumping, resulting from the formation of
CH$^+$ in excited rovibrational states following the reaction of C$^+$ with H$_2$, is
found to completely dominate the excitation of the vibrational transitions reported here.

5) The model is remarkably successful in explaining both the absolute and relative
strengths of the CH$^+$ and H$_2$ lines.

\begin{acknowledgements}

The observations reported here were carried out at the Infrared Telescope Facility (IRTF), 
which is operated by the University of Hawaii under contract NNH14CK55B with the 
National Aeronautics and Space Administration.  
We are very grateful to the IRTF director, John Rayner, for making unallocated engineering time available for the July 2019 observations that initiated this project.  We thank the IRTF support astronomers and telescope operators for the excellent support they provided for the
observations reported here.  
The grids of models have been run on the computing cluster Totoro funded by the European Research Council, under the European Community’s Seventh framework Programme, through the Advanced Grant MIST (FP7/2017-2022, No 742719).  M.G. is supported by the German Research Foundation (DFG) grant GO 1927/6-1.  P.B.C. is supported by NSF grant AST-1908576.
TRG’s research is supported by the international Gemini Observatory\footnote{A program of NSF's NOIRLab, which is managed by the Association of Universities for Research in Astronomy (AURA) under a cooperative agreement with the National Science Foundation, on behalf of the Gemini Observatory partnership: the National Science Foundation (United States), National Research Council (Canada), Agencia Nacional de Investigaci\'{o}n y Desarrollo (Chile), Ministerio de Ciencia, Tecnolog\'{i}a e Innovaci\'{o}n (Argentina), Minist\'{e}rio da Ci\^{e}ncia, Tecnologia, Inova\c{c}\~{o}es e Comunica\c{c}\~{o}es (Brazil), and Korea Astronomy and Space Science Institute (Republic of Korea).}. { We thank Holger M\"uller for useful discussions and the anonymous referee for several helpful suggestions.  Finally, we are particularly grateful to Miwa Goto, who expertly led the acquisition of the data at the IRTF.}

\end{acknowledgements}

\vfill\eject
\begin{appendix}
\section{Decomposition of blended spectral features}

\begin{figure}
\includegraphics[scale=0.8]{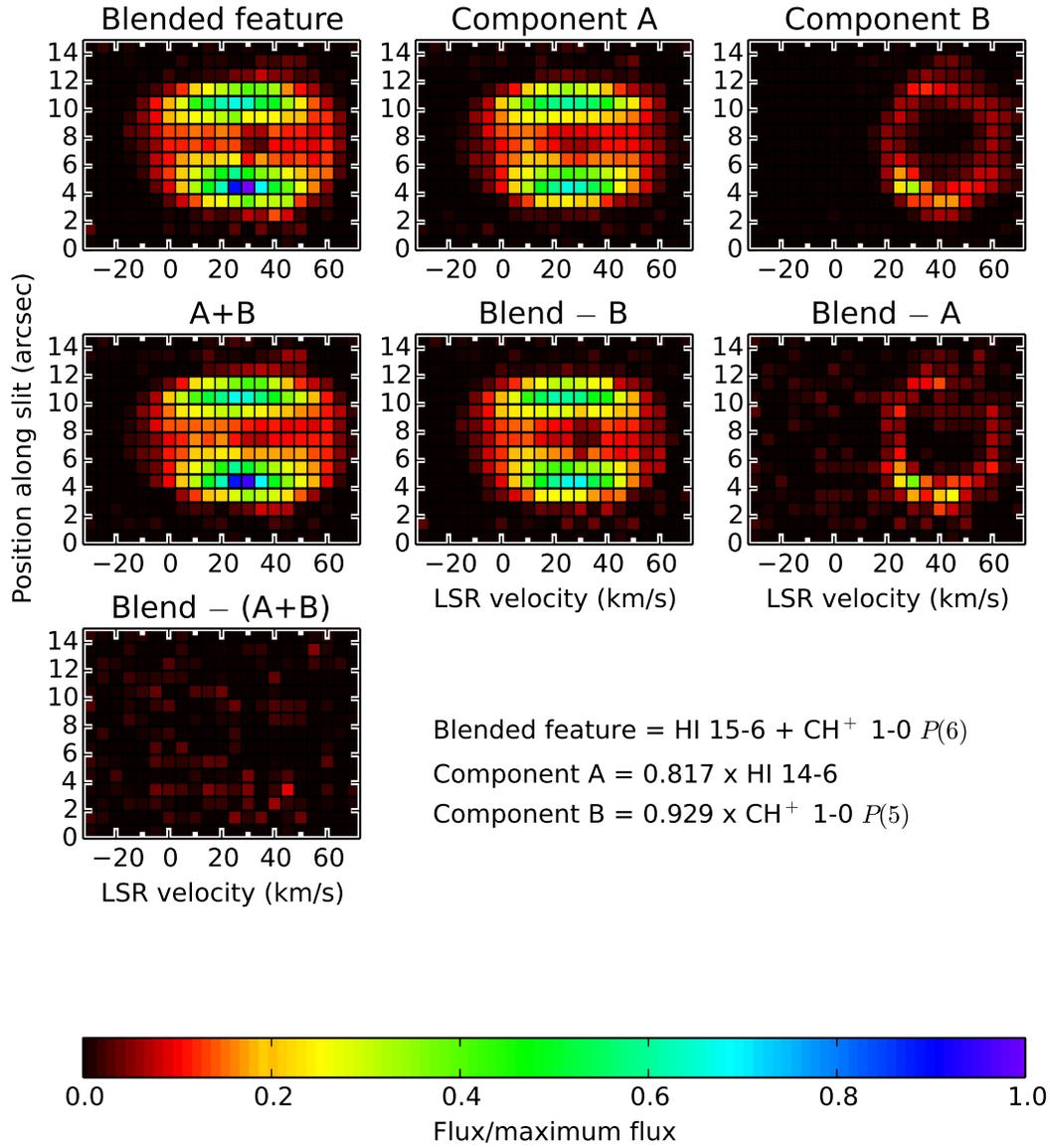}
\caption{Figure illustrating the decomposition method used to determine the flux of
the blended CH$^+$~$v=1-0$~$P(6)$ transition.  All {\it P-V} diagrams are computed for
the rest frequency of the HI $n=15-6$ transition, { and thus the LSR velocity refers to Component A.}  Details are described in the text.}
\label{fig:a1}
\end{figure}

The CH$^+$~$v=1-0$~$P(6)$ line at 3.907767$\,\mu$m is blended with the
nearby 3.907549$\,\mu$m $n=15-6$ recombination line of atomic hydrogen.  
To decompose the blended feature, we
fit the observed {\it P-V} diagram as a linear combination of those obtained for 
two unblended lines: those of the HI $n=14-6$ line and the CH$^+$~$v=1-0$~$P(5)$ 
line, with the latter shifted as appropriate to reflect  
difference between the CH$^+$~$v=1-0$~$P(6)$ and HI $n=15-6$ rest wavelengths.
The coefficients in the linear combination were adjusted to optimize the fit to the
data, yielding an estimated CH$^+$~$v=1-0$~$P(6)$/$P(5)$ line ratio of 0.927.
Figure~\ref{fig:a1} illustrates this deblending procedure. Here, the top row
shows {\it P-V} diagrams for the blended feature and 
the optimally-scaled HI $n=14-6$ CH$^+$~$v=1-0$~$P(5)$ lines (Components A and B).
The middle row shows the sum of Components A and B (left column), and the 
difference between the {\it P-V} diagrams for the blended feature and those for Components B (middle column)
and A (right column).  The optimal scaling we adopted therefore minimizes the differences 
between each {\it P-V} diagram in the middle row and the {\it P-V} diagram immediately above it.
The bottom left panel show the residuals in the fit, i.e. the difference between the
observed blended feature and the sum of Components A and B.

\begin{figure}
\includegraphics[scale=0.8]{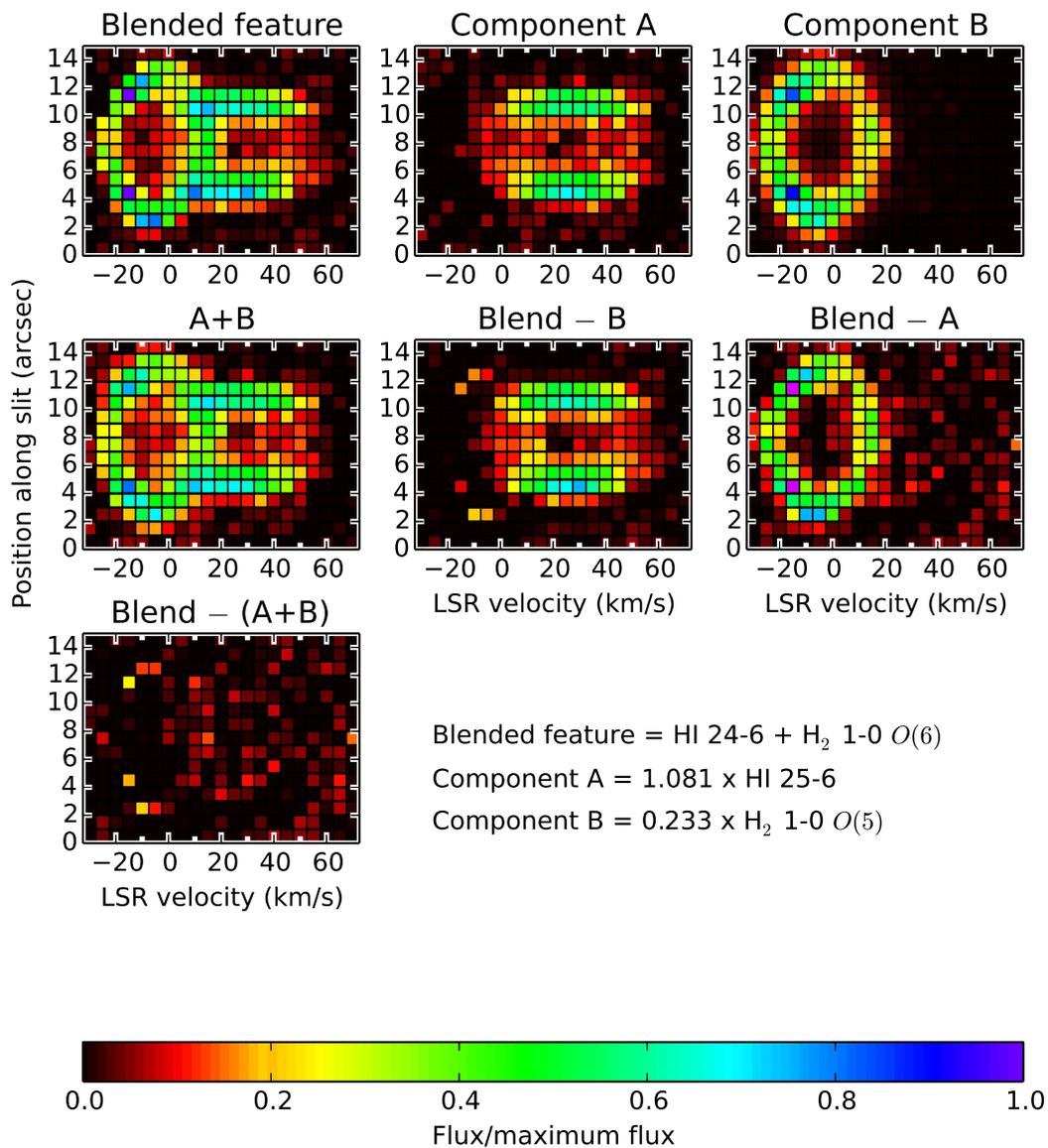}
\caption{Figure illustrating the decomposition method used to determine the flux of
the blended H$_2$~$v=1-0$~$O(6)$ transition.  All {\it P-V} diagrams are computed for
the rest frequency of the HI $n=25-6$ transition,  { and thus the LSR velocity refers to Component A.} Details are described in the text.}
\label{fig:a2}
\end{figure}

The H$_2$~$v=1-0$~$O(6)$ line at 3.500809$\,\mu$m is blended with the
nearby 3.501164$\,\mu $m $n=24-6$ recombination line of atomic hydrogen.
We adopted a similar deblending procedure to that described above, but now using a linear 
combination of the PV diagrams for H$_2$~$v=1-0$~$O(5)$ and HI $n=25-6$.  This procedure
is illustrated in Figure~\ref{fig:a2}. 

\clearpage

\section{Collisional excitation of CH$^+$}

\subsection{Nonreactive collisional excitation}

For \bl{the collisional excitation of CH$^+$ within the ground vibrational
state by e$^-$ and H, we adopt the data of Hamilton et al.\ (2016) and Faure et al.\ (2017), who provided deexcitation rates for states up to $J_{\rm max}=18$} (Faure, private communication). For collisions with He, we use the deexcitation rates computed by Hammami et al.\ (2009) \bl{for states up to $J_{\rm max}=10$}. Collisional rates for \CHp\ with \HH\ are derived from those of \CHp\ with He using the rigid rotor approximation.

The above deexcitation rate coefficients are used for any transition $v',J' \rightarrow v'',J''$ such that $v'=v''=0$ and $J' \leqslant J_{\rm max}$. The deexcitation rate coefficient $k^{\rm C}_{v'J'v''J''}$ of other rotational or vibrational transition is computed as 
\begin{equation}
k^{\rm C}_{v'J'v''J''} = k^{\rm C}(T)  \, f_{v'J'v''J''},
\end{equation}
where $k^{\rm C}(T)$ is the total deexcitation rate coefficient, which we assume identical for all levels $v',J'$ and consider to only depend on the kinetic temperature $T$, and $f_{v'J'v''J''}$ is a distribution function over lower levels such that
\begin{equation}
\sum_{v'',J''} f_{v'J'v''J''} = 1 \quad \forall \,\, v',J'.
\end{equation}
The total deexcitation rate coefficient $k^{\rm C}(T)$ is obtained by performing a second order polynomial fit in \bl{logarithmic space to} the total deexcitation rates for pure rotational levels,
\begin{equation} \label{Eq-total-rate}
k^{\rm C}(T) = 10^{\alpha} \, 10^{\beta \, {\rm log}(T)} \, 10^{\gamma \, {\rm log}^2(T)}.
\end{equation}
\bl{We assume that the} distribution function $f_{v'J'v''J''}$ \bl{can be approximated as}
\begin{equation} \label{Eq-proba-dist}
f_{v'J'v''J''} = a(\Delta v) \, b(\Delta J) \,(2J''+1) \, \Delta E^c,
\end{equation}
i.e.\ \bl{follows a power dependence} on the level energy difference $\Delta E$ with scaling coefficients $a$ and $b$ that solely depend on $\Delta v = v' - v''$ and $\Delta J = J' - J''$, respectively. The power-law dependence on $\Delta E$ is deduced from a fit \bl{to} the pure rotational collisional rates. The $b$ coefficients are set to reproduce the dependence on $\Delta J$ of pure rotational transition for $\Delta J \leqslant \Delta J_{\rm max}$ and set to 0 for $\Delta J > \Delta J_{\rm max}$.  Here, $\Delta J_{\rm max} = 18$, 8 and 10, respectively, for collisions with H, e$^-$, and He.
The $a$ coefficients are set to 1 for $\Delta v = 0$ and to 0.1 for $\Delta v \neq 0$ in order to reproduce the  vibrational deexcitation rates of the $v=1,2,3$ levels of \CHp\ by collisions with e$^-$ recently computed by Jiang et al.\ (2019).

\begin{figure}[!ht]
\begin{center}
\includegraphics[width=12cm,trim = 1cm 2.0cm 0cm 2.0cm, clip,angle=0]{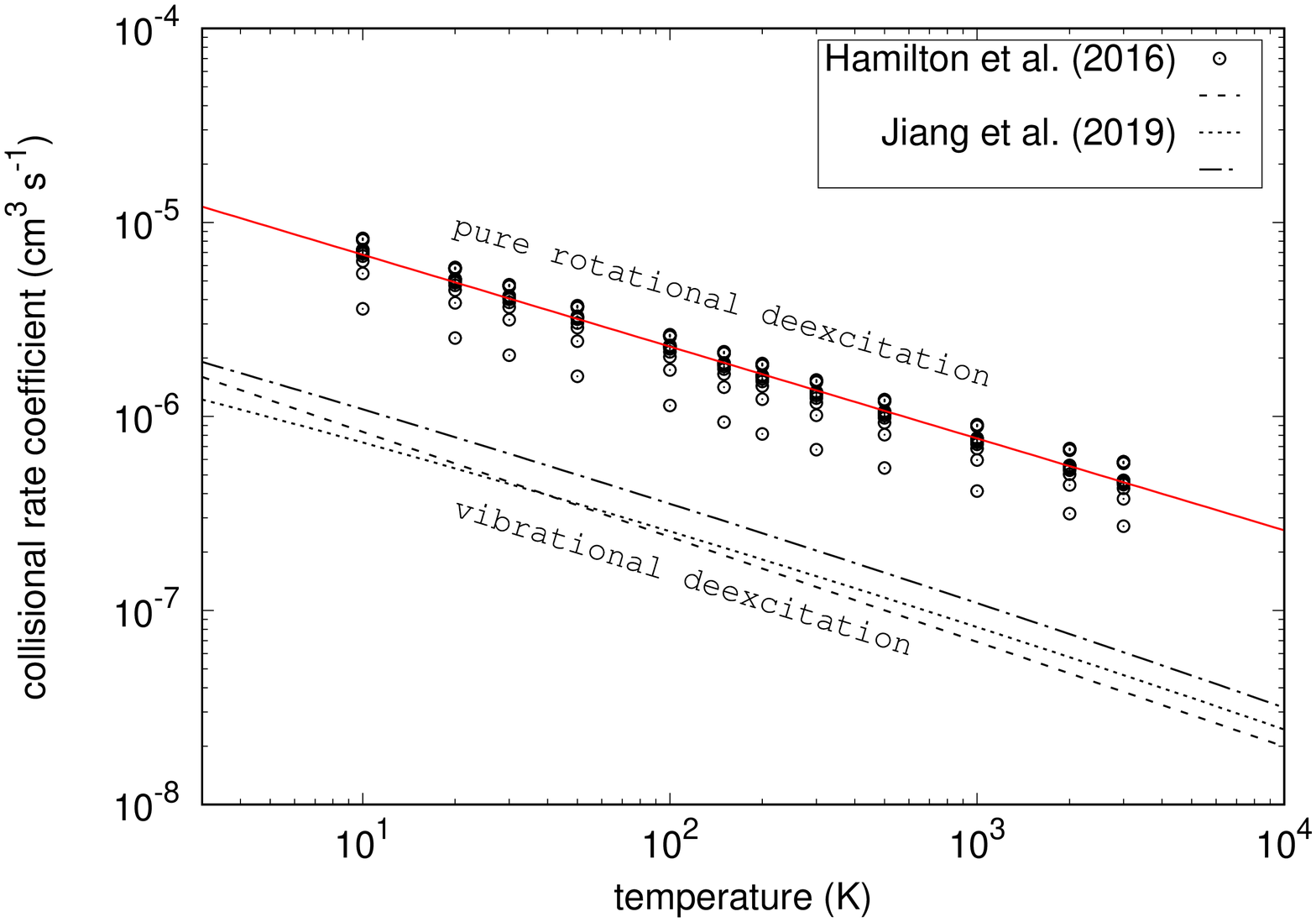}
\includegraphics[width=12cm,trim = 1cm 2.0cm 0cm 2.0cm, clip,angle=0]{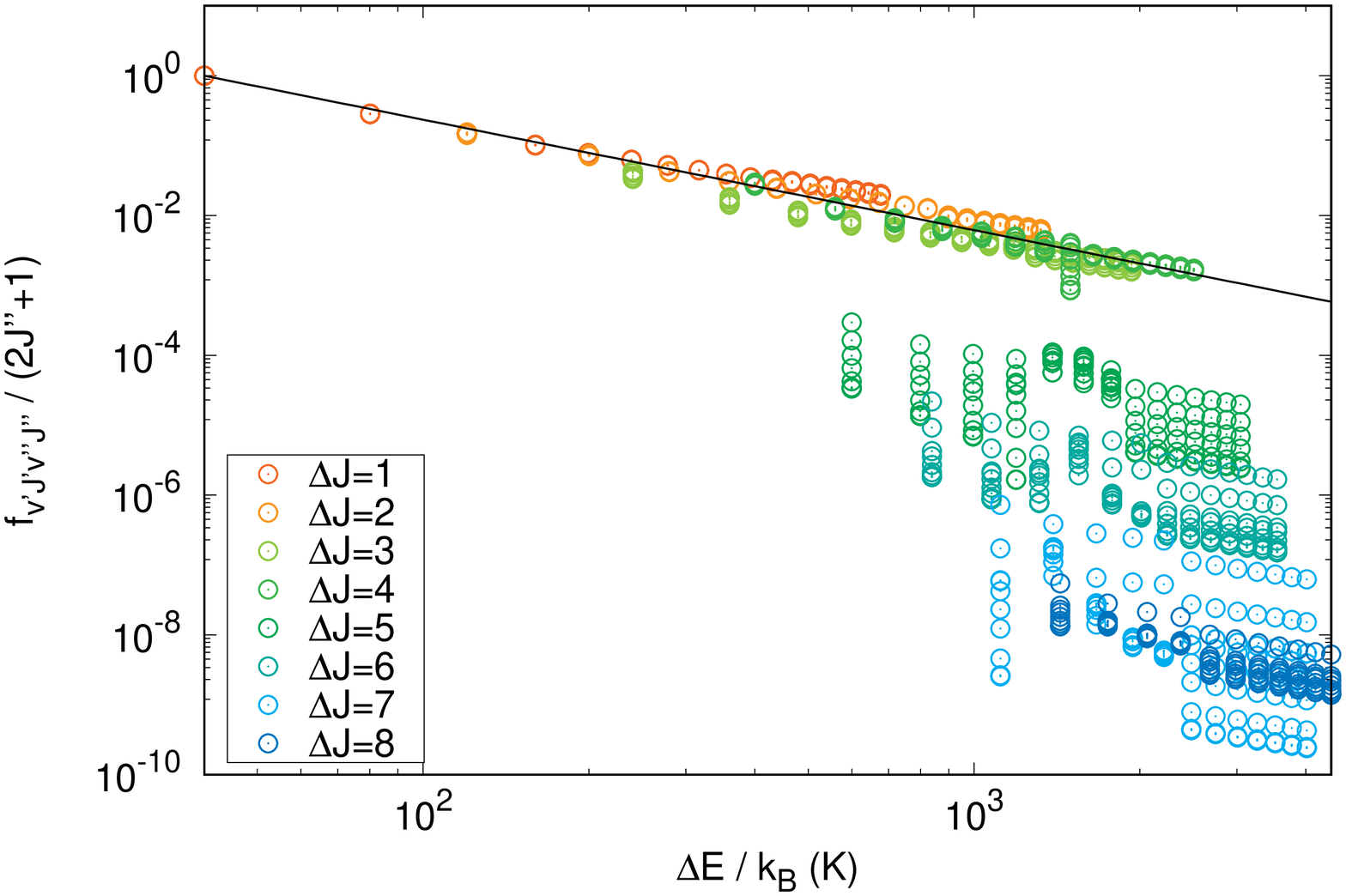}
\caption{{\it Top}: total collisional deexcitation rate coefficients for \CHp\ in collisions with e$^-$. Open circles correspond to deexcitation of pure rotational levels (Hamilton et al.\ 2016). Black curves correspond to vibrational deexcitation of the $v'=1$ (dashed), $v'=2$ (dotted), and $v'=3$ (dotted-dashed) levels of \CHp\ (Jiang et al.\ 2019). The red curve shows a fit of pure rotational data using Eq. \ref{Eq-total-rate} with $\alpha = -4.7$, $\beta =-0.47$, and $\gamma = 0$. {\it Bottom}: probability distribution of rotational deexcitation as function of the level energy difference $\Delta E$ and $\Delta J$ (Hamilton et al.\ 2016). The black curve indicates the power-law fit used in Eq. \ref{Eq-proba-dist}.}
\label{Fig-coll-rates}
\end{center}
\end{figure}

An illustration of the prescription adopted for the \CHp -e$^-$ collisional system is given in Figure \ref{Fig-coll-rates}. Although empirical, the above treatment has the advantage of limiting the collisional rates to the total deexcitation rate of pure rotational levels and to provide simple and separated prescriptions of the dependences on $\Delta E$, $\Delta v$, and $\Delta J$ that favor elements of the collisional matrix close to the diagonal relative to those far from the diagonal, in fair agreement with the available data.

\subsection{Reactive collisional excitation}

Adopting the prescription of the Meudon PDR code, the total reaction rates of reaction \ref{Eq-react} are first computed using the quasi-classical treatment of Herr{\'a}ez-Aguilar (2014), who derived rate constants for ten rovibrational levels of \HH\ in $v=0$ and $v=1$.  \bl{They are then scaled} to reproduce the detailed time-dependent quantum calculations of Zanchet et al.\ (2013, hereafter Z13), who derived rate constants for the $(v,J)=(0,0)$, $(1,0)$, and $(1,1)$ levels of \HH. The total reaction rates for all \HH\ levels above $(v,J)=(2,0)$ are set to the value obtained by Faure et al.\ (2017, hereafter F17) for that level.

\begin{figure}[!ht]
\begin{center}
\includegraphics[width=14 cm,trim = 1cm 2.0cm 0cm 1.0cm, clip,angle=0]{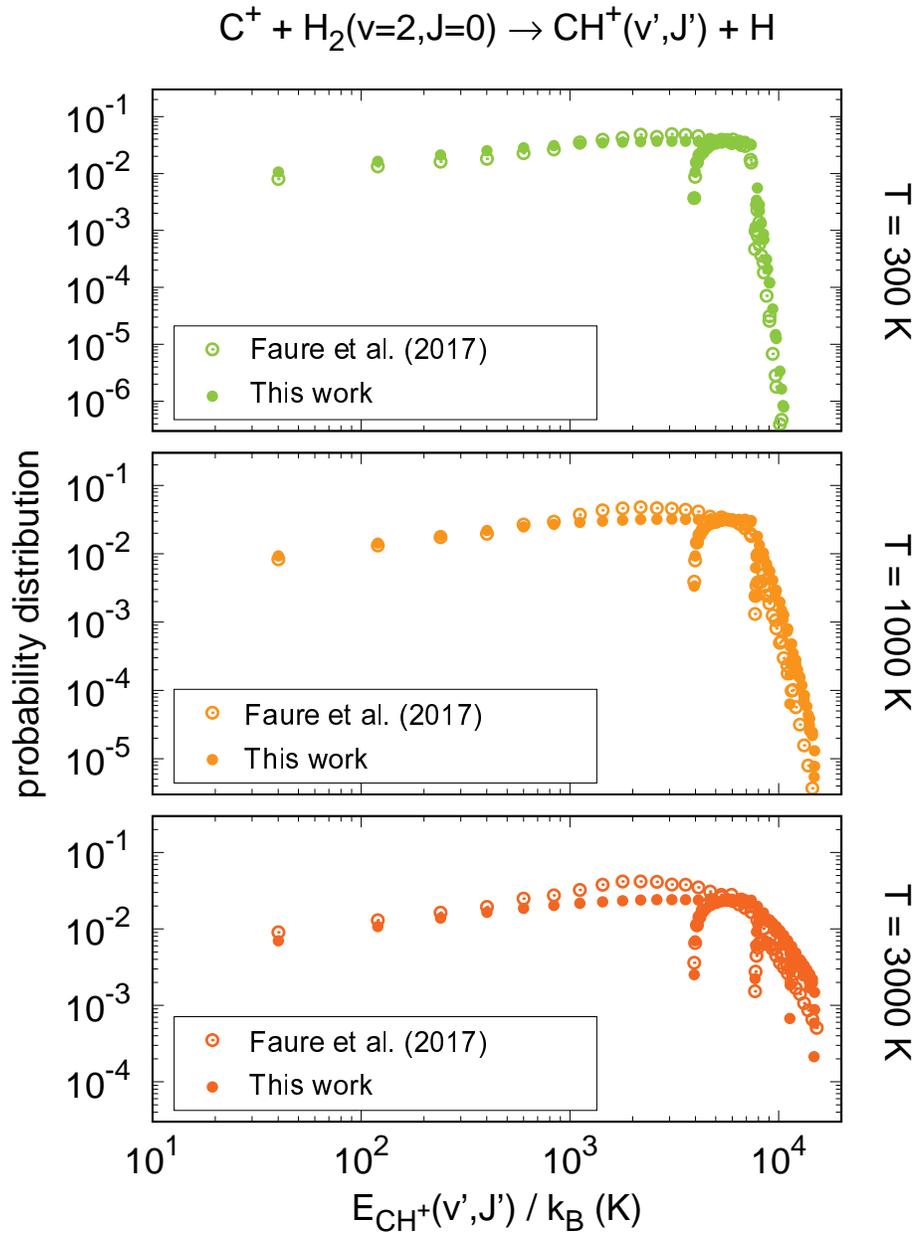}
\caption{Probability distributions of forming \CHp\ in its rovibrational levels $(v',J')$ through the reaction of $\Cp$ with $\HH(v=2,J=0)$ as function of the level energy $E_{\CHp}(v',J')$. The distributions computed with equation \ref{Eq-chem-proba-distr} (filled circles) are compared with those obtained by F17 (empty circles) for a kinetic temperature of 300 K (top panel), 1000 K (middle panel), and 3000 K (bottom panel).}
\label{Fig-chem-distr}
\end{center}
\end{figure}

The probability distribution of forming \CHp\ in its rovibrational levels is deduced from the recent studies performed by Z13 and F17. For any level $(v,J)$ of \HH, the probability $p(v',J')$ of forming \CHp\ in a level $(v',J')$ is assumed to scale as
\begin{eqnarray} \label{Eq-chem-proba-distr}
p(v',J') \propto &(2J'+1) \, {\rm exp}(-0.08 J') \phantom{ \, {\rm exp}(-E/k_B T)}        &\rm{if} \,\,\,E <         0 \\
&(2J'+1) \, {\rm exp}(-0.08 J') \, {\rm exp}(-E/k_B T)  &\rm{if} \,\,\,E \geqslant 0, \nonumber
\end{eqnarray}
where
\begin{equation}
E = E_0 + E_{\CHp}(v',J') - E_{\HH}(v,J)
\end{equation}
is the total energy balance \bl{for} reaction \ref{Eq-react}, including the enthalpy $E_0$ of the reaction ($E_0 /k_B = 4281$ K) and the energies $E_{\CHp}(v',J')$ and $E_{\HH}(v,J)$ of the $(v',J')$ and $(v,J)$ rovibrational levels of \CHp\ and \HH, respectively\footnote{In our treatment of reactive collisional excitation, we neglect any dependence of $p(v',J')$ on the rotational state, $J$, of the reactant H$_2$ molecule.  
To date, no quantum calculations are available for any case with $J > 1$; future calculations would be extremely desirable to determine the dependence on $J$, if any.}.    
Examples of the probability distributions obtained with Equation \ref{Eq-chem-proba-distr} for $(v,J)=(2,0)$ are shown in Figure \ref{Fig-chem-distr} and compared with the distributions calculated by F17. Although the above recipe appears to overestimate the production of \CHp\ in highly excited states at high kinetic temperature, it provides a simple prescription that captures relatively well the strong dependences of the occupation probabilities on the total energy balance of reaction \ref{Eq-react} and on the kinetic temperature.

\end{appendix}

\end{document}